\documentclass[english,11pt,oneside]{article}

\pdfoutput=1
\usepackage[T1]{fontenc}
\usepackage[latin9]{inputenc}
\usepackage[usenames,dvipsnames]{xcolor}
\usepackage{color}
\usepackage{babel}
\usepackage{setspace}
\usepackage{amstext}
\usepackage{amssymb}
\PassOptionsToPackage{normalem}{ulem}
\usepackage{ulem}
\usepackage[unicode=true,pdfusetitle,
 bookmarks=true,bookmarksnumbered=false,bookmarksopen=false,citecolor=Turquoise,
 breaklinks=false,pdfborder={0 0 1},backref=false,colorlinks=true,pdfpagemode=FullScreen,linktoc=page]
 {hyperref}
\usepackage{graphicx}
\usepackage{mathtools}
\usepackage[small]{caption}
\usepackage{changepage}
\usepackage{slashed}
\providecommand{\tabularnewline}{\\}
\definecolor{note_fontcolor}{rgb}{0.80078125, 0.80078125, 0.80078125}
\usepackage[top=3 cm, bottom=3 cm, left=3.1416 cm, right=3.1416 cm]{geometry}

\newcommand{\ka}[1]{\textbf{\textcolor{blue}{(#1 --KA)}}}

\def\beq{\begin{equation}}
\def\eeq{\end{equation}}
\def\bea{\begin{eqnarray}}
\def\eea{\end{eqnarray}}
\newcommand{\ba}{\begin{eqnarray}}
\newcommand{\ea}{\end{eqnarray}}
\newcommand{\no}{\nonumber}

\begin{document} 

\baselineskip=17pt

%%%%%%%%%%
%%%%%%%%%%    Title page
%%%%%%%%%%

\thispagestyle{empty}
\vspace{20pt}
\font\cmss=cmss10 \font\cmsss=cmss10 at 7pt

\begin{flushright}
\today \\
UMD-PP-015-017\\
\end{flushright}

\hfill
%\vspace{20pt}

\begin{center}
{\Large \textbf
{Warped Seesaw is Physically Inverted}}
\end{center}

\vspace{15pt}

\begin{center}
{\large Kaustubh Agashe$\, ^{a}$, Sungwoo Hong $\, ^{a}$, Luca Vecchi $\, ^{a, b}$ \\
\vspace{15pt}
$^{a}$\textit{Maryland Center for Fundamental Physics,
     Department of Physics,
     University of Maryland,
     College Park, MD 20742, U.~S.~A.} \\
      $^{b}$\textit{Dipartimento di Fisica e Astronomia, Universita` di Padova, and INFN Sezione di Padova, Italy} 
      \\ 
{\it email addresses}: kagashe@umd.edu; sungwoo83hong@gmail.com; vecchi@pd.infn.it

}

\end{center}

\vspace{5pt}

\begin{center}
\textbf{Abstract}
\end{center}
\vspace{5pt} {\small \noindent

Warped extra dimensions can address both the Planck-weak and flavor hierarchies of the Standard Model (SM). In this paper we discuss the SM neutrino mass generation in a scenario in which a SM singlet bulk fermion --- coupled to the Higgs and the lepton doublet near the IR brane --- is given a {\em Majorana} mass of order the Planck scale on the UV brane. Despite the resemblance to a type I seesaw mechanism, a careful investigation based on the {\em mass} basis for the singlet 4D modes reveals a very different picture. Namely, the SM neutrino masses are generated dominantly by the exchange of the {\em TeV}-scale mass eigenstates of the singlet, that are pseudo-Dirac and have a sizable Higgs-induced mixing with the SM doublet neutrino: remarkably, in warped 5D models the anticipated type I seesaw morphs into a {\em natural} realization of the so-called ``inverse'' seesaw. This understanding uncovers an intriguing and direct link between neutrino mass generation (and possibly leptogenesis) and {\em TeV}-scale physics. We also perform estimates using the dual CFT picture of our framework, which back-up our 5D calculation.

%{\color{blue} The phenomenological implications of our above finding can then be potentially striking, for example, the physics of the SM neutrino mass generation (namely, the properties of the TeV mass singlet modes) can be probed {\em directly} at the LHC/future colliders. In addition, leptogenesis can proceed via decays of these new particles, even if, in this 5D model, the universe is reheated to only $\sim$ TeV temperatures. {\bf Is this true?} Clearly, both these tasks are not possible in {\em high}-scale type I seesaw models.}

}
 
\vfill\eject
\noindent

%%%%%%%%%%
%%%%%%%%%%    Main Text
%%%%%%%%%%

\section{Motivation and summary}
\label{intro}

The Randall-Sundrum (RS1) model \cite{Randall:1999ee}
with a warped extra dimension [in particular, five-dimensional (5D) anti de-Sitter space (AdS)], coupled with an appropriate mechanism 
\cite{Goldberger:1999uk} to stabilize 
the size of the extra dimension, provides an attractive solution to
%
%addresses 
%
the Planck-weak hierarchy problem
of the Standard Model (SM). The basic idea is that localizing the SM Higgs boson near the IR brane results in
scale of its vacuum expectation value (VEV) being warped-down to the $\sim$ TeV scale relative to that of 4D graviton (i.e., the Planck scale) 
which is localized near the UV brane.
By the correspondence between AdS space and conformal field theories (CFTs) in lower space-time dimension \cite{Aharony:1999ti}, this 
idea
%
%framework 
%
is dual to a purely 4D theory, where the SM Higgs boson is a composite of some
%
%(unkown/unspecified) 
%
new strong dynamics \cite{ArkaniHamed:2000ds}.

In addition, the warped framework with the SM fermions arising as zero-modes of fermion fields propagating in the extra dimension can also
%
%address
%
account for the {\em charged} fermion mass and mixing angle (flavor) hierarchies of the SM as follows 
\cite{Grossman:1999ra, Gherghetta:2000qt, Huber:2000ie}. 
%
%Namely, 
%
The effective 4D Yukawa couplings are dictated by the {\em overlap} of fermion zero-mode profiles with the
Higgs boson, the latter being localized near/on the TeV/IR brane.
%
%In turn, 
%
The crux of this idea is that 
small changes in the {\em five}-dimensional (5D) mass parameters can result in large variations in the  
(extra-dimensional) profiles of the fermion zero modes at the TeV brane, 
thus 
(easily) generating the desired hierarchies in these Yukawa couplings, i.e., the SM fermion masses.
%
%Remarkably
%
It is interesting
%
%fascinating 
%
that such a scenario for SM fermions is 
dual to  
SM fermions being partially composite also \cite{Contino:2004vy}, to degrees determined by scaling dimensions of the
%
%(different) 
%
fermionic operators to which they couple (this scaling dimension is
%
%being 
%
dual to the 5D mass parameter).
The point then is that the 
coupling to Higgs is dictated by the amount of composite admixture in SM fermions, which can be hierarchical even with small 
%
%variations/changes 
%
differences
in the scaling dimensions of the fermionic operators, provided there is a large energy range for the associated 
renormalization group evolution (RGE).
%
%RG scaling.
%
Of course, 
5D fermions 
necessitate 5D gauge fields \cite{Davoudiasl:1999tf}.

In such a ``bulk'' SM in warped extra dimension (see also \cite{Goldberger:1999wh}), there are also Kaluza-Klein (KK) excitations of SM particles, which have masses starting at and quantized in units of roughly TeV scale and profiles which are peaked
near the TeV brane. These 
%
%beyond the SM
%
new particles inherently contribute to various types of precision tests of the SM. Thus, there are indirect constraints on the KK mass scale in this model; 
the worry being that KK scale much larger than $\sim$ TeV will jeopardize the solution to the Planck-weak hierarchy problem.
Those from 
electroweak tests can be controlled by suitable 
custodial symmetries \cite{Agashe:2003zs}, allowing a few TeV KK scale \cite{Carena:2006bn}.
As far as flavor violation  is concerned, there is a 
built-in suppression of such effects in this framework, roughly an analog of Glashow-Iliopoulos-Maiani (GIM) mechanism in 
the SM \cite{Gherghetta:2000qt, Huber:2000ie, Agashe:2004cp}.
%
%In spite of it/
%
Still, KK scale above $\sim 10$ TeV might be required (modulo the option of fine-tuning of flavor parameters) 
in order to be consistent with flavor precision data \cite{Csaki:2008zd}.
Of course, this situation can be 
mitigated by use of appropriate flavor symmetries \cite{Barbieri:2012tu} such that a few TeV KK mass scale can be once again allowed\footnote{In addition, there are lower bounds on the KK scale from absence of any signal of {\em direct} production of these
KK particles at the LHC, but those from run 1 are still below the few TeV limit that we get from precision tests.}.
For a review of the framework and its phenomenology (and more references), see, for example, \cite{Davoudiasl:2009cd}.

In this paper, we study the SM {\em neutrino} masses in this framework:
%
%Moving onto SM neutrino masses, 
%
clearly there are two options to begin with, namely, Dirac or Majorana type mass. 
For Majorana neutrinos, an incarnation of the  
%
%usual
%
standard type I seesaw mechanism \cite{seesaw} has been incorporated in the warped extra dimensional framework
\cite{Huber:2003sf, Csaki:2003sh, Perez:2008ee}: we will focus only on this model in this paper.\footnote{For other scenarios (for either Dirac or Majorana case) see, for example, references \cite{Grossman:1999ra, Gherghetta:2003he, Fong:2011xh}.}
In this model, SM singlet neutrinos are added in the bulk
to the above framework of SM-charged fermions, aka the ``right-handed'' (RH) neutrino in the
4D case, 
even though it gives
massive 
4D modes with {\em both} chiralities in the 5D version (a fact which will turn out to be crucial in our work). This singlet 
%
%RH 
%
neutrino field has a coupling to lepton doublet 
%
%($L$) 
%
and 
Higgs on (or near) the IR brane, from
%
%via 
%
which the 
%
%would-be zero mode of 
%
%RH
%
singlet neutrino 
5D 
field 
acquires a Dirac  mass term with the doublet (or LH) neutrino field 
once EW symmetry breaking (EWSB) occurs, i.e., Higgs develops a VEV
%
%zero mode
%
(just like for charged SM fermions).
However, the {\em difference} from charged fermion case is that we assume that 
lepton-number is broken only on the UV brane (i.e., it is still a good symmetry in the bulk and on the TeV brane).
This choice essentially 
manifests itself as a Majorana mass term for the UV brane-localized value of the bulk singlet neutrino field.
(Obviously, no such mass terms are allowed for the charged fermions.)
%
%

%
%note on bulk extended EW gauge symmetry forbidding Mjorana mass elsewhere...
%
Note that adding a Majorana mass term (or lepton-number violation) only on the UV brane is technically natural by 5D locality. It is also quite generic in scenarios where the 
bulk EW gauge group is extended to $SU(2)_L \times SU(2)_R \times U(1)_{B-L}$ in order to satisfy bounds from EW precision tests \cite{Agashe:2003zs}. Here $SU(2)_R \times U(1)_{B-L}$ is spontaneously broken down to $U(1)_Y$ (hypercharge of the SM) on the Planck brane, either by boundary conditions or
Planckian VEV of a localized scalar (this is equivalent to the former case in the large VEV limit), whereas $SU(2)_L \times U(1)_Y \rightarrow U(1)_{ \rm EM }$ occurs by the Higgs VEV localized near the IR brane. In this setup $N$ will be typically charged under $SU(2)_R \times U(1)_{B-L}$\footnote{In fact, in the canonical case, this SM singlet simply corresponds to the $SU(2)_R$ doublet partner of
the charged RH lepton, i.e., it is not added ``by hand'', rather its presence is required by the bulk gauge symmetry.} while remaining neutral under the SM gauge group.
Such a choice of the bulk gauge symmetry (and breaking)
implies that a Majorana mass term for $N$, which would break $SU(2)_R \times U(1)_{B-L}$, is only
allowed on Planck brane, i.e., it is forbidden in the bulk and on TeV brane.

We contextualize our contribution by first recapitulating the approaches used in previous studies.
%
%We will first briefly review previous results for this model and then outline our new work.
%
It turns out that 
%
%\begin{itemize}
%\item
most of the earlier studies of this model  \cite{Huber:2003sf, Perez:2008ee} were performed 
%
%using
%
employing the ``usual'' (i.e., similarly to the {\em charged}-SM fermions) KK modes of the SM {\em singlet} field 
as the basis, where the above-mentioned 
Planck brane localized 
Majorana mass term is treated as a (not necessarily small) ``perturbation'' or at the least an ``add-on'': we will call this simply the ``KK'' basis.~\footnote{An exception is reference \cite{Csaki:2003sh}, which employed the {\em full} mass basis, i.e., for all modes (entire tower) of neutrinos (i.e., diagonalizing {\em also} the effect of doublet and singlet mixing due to EWSB, which we neglect here to begin with, rather it can be 
 genuinely treated as 
a insertion/perturbation).
However, this study 
%
%mostly
% 
focussed only on mass of the {\em lightest} (i.e., mostly SM) neutrino state, i.e., it did {\em not} (at least explicitly) work out the 
spectrum of  heavier states. Hence, the ``inner workings'' of the SM neutrino mass, whose exchange is responsible for its generation,
%
%anatomy 
%
is 
not clear from such an analysis.}

%\end{itemize}
%
In more detail, in these earlier papers the KK decomposition for {\em singlet} field \footnote{At leading order in Higgs VEV, the {\em doublet} 
lepton {\em KK} modes will play no role in the generation of the SM neutrino mass, no matter which basis we use. So, we will only
keep the doublet {\em zero}-mode, i.e., (approximately) the SM doublet lepton, from now on.} 
is performed {\em neglecting} the Majorana mass on UV brane, giving zero (chiral) and
massive, Dirac (KK) modes, just like for doublet lepton and, in general, SM charged fermion fields.
Afterwards, turning on the Planck brane localized Majorana mass term
results in the would-be zero-mode acquiring a large Majorana mass. Furthermore, it leads to mixing (via Majorana mass terms) {\em among} the would-be zero {\em and} (already massive) KK modes so that clearly the would-be zero modes and KK modes are {\em not} the mass eigenstates. %(again, this KK basis {\em would} have been the mass basis -- of course {\em before} EWSB -- had there been {\em no} Majorana mass term for singlet field on the UV brane, as is the case for SM-charged fields).
%
%these zero and KK modes would have been mass eigenstates...
%
%Turning on 
%
Finally, including 
EWSB leads to mass terms between the SM neutrino 
and the entire tower of singlet modes; integrating out the latter then 
generates a mass for the SM neutrino, which is thus purely Majorana in nature,
deriving from the above-mentioned Majorana mass terms for the singlet modes.
The {\em advantages} of the KK basis are its familiarity (from the numerous studies of {\em charged} fermion masses, where of course 
such Majorana mass terms are absent). As we will detail in what follows, it is perhaps the 
quickest/easiest way to obtain the SM neutrino mass formula in the 5D model. Indeed, 
%
%\begin{itemize}
%
%\item
the exchange of non-zero {KK} singlet modes with Dirac mass terms quantized in units of TeV-scale 
gives {\em negligible} contribution to the SM neutrino mass ({\em inspite} of these modes having Majorana mass terms also):
%
%whereas 
%
almost all of this effect then comes instead from the 
%
%exchange of 
%
would-be {\em zero}-mode (i.e., {\em no} Dirac mass term), with 
a super-large Majorana mass term. 
This ``anatomy'' of the SM neutrino mass 
gives it the appearance of {\em type I high-scale} seesaw. 

%\end{itemize}
%
In addition, the ``intermediate'' seesaw scale which is typically needed in type I high-scale seesaw models for obtaining the right SM neutrino mass 
can be naturally realized in the 5D model, i.e., even with input parameters being Planckian, 
via a natural choice of 5D mass of the singlet. In contrast, in 4D models such a seesaw scale often has to be introduced as a ``new'' scale.

In this paper, we {\em re}-consider the model using the {\em mass} basis (instead of the above KK one) for the singlet
4D modes, neglecting the mass mixing with doublet due to Higgs VEV. The reason is that this is the basis necessary to analyze processes involving {\em on}-shell singlet neutrinos, such as direct collider signals of singlet neutrino states and leptogenesis \cite{future}.
%\end{itemize}
%

What we find is that the character of the seesaw is ``changed'' when the mass basis is employed! Namely, even though the SM neutrino mass is obtained exchanging the mass eigenstates of the singlet (similarly to exchanging would-be KK modes), we show that

\begin{itemize}

\item
%
%namely, 
%
the {\em TeV}-scale {\em mass} eigenstates of the singlet actually give a {\em significant} contribution to the SM neutrino mass
(the end result being of course the same as in KK basis); in fact, this is the {\em dominant} effect for the natural versions of the model.
\end{itemize}
Given {\em also} their {\em un}suppressed Yukawa couplings to Higgs and the 
SM neutrino (following from their profile leaning towards TeV brane, where Higgs is also localized), 
at first sight, it seems somewhat counter-intuitive that 
the SM neutrino mass comes out very small: indeed, this is 
%
%In spite of TeV mass and couplings 
%
due to 
these modes being mostly Dirac, i.e., with a highly suppressed Majorana mass term.

A similar mechanism in {\em four} dimensions goes by the name ``inverse'' seesaw  \cite{Mohapatra:1986aw}, i.e., where the very small SM neutrino mass
arises from  exchange of (possibly TeV-mass) singlet mode which is pseudo Dirac and
has sizable EWSB mass term with the SM neutrino.
%
%i.e., with a small Majorana mass term, 
%
Thus, we discover that, in mass basis, the dynamical picture of a seemingly high-scale Type-I seesaw model in warped 5D is that of an
``inverse'' see-saw.
Actually, it is crucial that 
%
%\begin{itemize}
%\item
the Majorana mass term for these TeV-mass modes in the 5D model 
is {\em naturally} small, as opposed to generic 4D inverse seesaw models, where such a smallness can be rather an ad-hoc assumption.
%\end{itemize}
%

Phenomenologically, we then see that -- for the purpose of 
leptogenesis or probing {\em directly} the mechanism of the SM neutrino mass generation in this 5D model by
producing the responsible singlet states at the LHC/future colliders -- the center of attention becomes {TeV}-mass
singlet modes, as in the usual/4D inverse seesaw
models. %cf.~(much) heavier scales that one would have expected to deal with based on the KK basis impression of type I (high-scale) seesaw.

%\end{itemize}
%

Furthermore, the  CFT interpretation of this seesaw 
%
%neutrino mass 
%
model has not been discussed in the literature thus far,
even though the charged SM fermion case has been thoroughly studied in this way, providing physical intuition to the problem. % and back-of-the-envelope (but still reliable!)
Such a dual CFT description of warped seesaw for neutrino masses 
will be similarly extremely useful, offering an alternative picture for SM neutrino mass generation.
In fact, we find that 
\begin{itemize}

\item
%
%The 
%
%advantage 
%
%chief utility of the CFT viewpoint is that 
%
the CFT viewpoint allows us to quickly unveil the true nature of the seesaw mechanism and clarifies the naturalness of the small Majorana masses of TeV-scale eigenstates. 
%
%{\em both} of which 
%
%these features are seen only upon a tedious calculation in the mass basis of the 5D model!

%
\end{itemize}
%

%In addition, it is rather straightforward to {\em estimate} the SM neutrino mass in the CFT basis: one can then use the standard AdS/CFT dictionary for these models in order to translate the SM neutrino mass expression to the one involving the corresponding 5D model parameters, thus providing a check against those calculations.

%%%%%%%%%%%%%%%%%%%%%%%%%%%%%%%%%%%%%%%
% TABLE
%%%%%%%%%%%%%%%%%%%%%%%%%%%%%%%%%%%%%%%
\begin{table}[tbp]
\begin{adjustwidth}{-1cm}{}
%\centering
\scalebox{0.65}
{
\begin{tabular}{ | c || c || c | c | }
\hline 
Basis $\rightarrow$ & KK & mass & CFT  \\
Features  &  (would-be mass modes {\em neglecting}  & (for singlet only, i.e., &  [$N_R$ (external) and \\
$\downarrow$ &  UV brane Majorana mass term) & neglecting Higgs VEV) & composites (with 2 sectors {\em mixing})]  \tabularnewline
\hline
\hline
Advantage/Use & familiar from charged fermion analysis;  & needed for {\em on}-shell production & elucidates seesaw structure \tabularnewline
& easy to obtain $m_{ \nu }$ & (LHC and/or leptogenesis) & easy to obtain $m_{ \nu }$ \tabularnewline
& & & ``bridge'' between mass and KK bases \tabularnewline 
\hline
Nature of seesaw & Type I (high-scale) & (Dominantly) inverse for $c_N > -1/2$ & (Significantly) inverse \tabularnewline
(details below) & (for {\em both} 
$c_N <-1/2$ and $>-1/2$)
& ``Combination'' for $c_N < -1/2$ &  (for {\em both} 
%
%$c_N >1/2$ and $<1/2$
%
$\big[ {\cal O}_N \big] > 5/2$ and $< 5/2$) 
\tabularnewline
\hline
fraction of (net) $m_{ \nu }$ & $0$ &  $\approx  1$ ($\sim 1$) for $c_N >  (<)-1/2$ & $\sim1$ (for {\em both} 
%
%$c_N >1/2$ and $<1/2$
%
$\big[ {\cal O}_N \big] > 5/2$ and $< 5/2$) \tabularnewline
from $\sim$ TeV-scale modes & (from each {\em Dirac} 
%
%state/
%
mode) & (from pseudo-Dirac {\em pair}) &  (from each Dirac composite) 
\tabularnewline
\hline
%
%Super-
%
heavy (Majorana) mode
%
%mode 
%
& would-be zero-mode, {\em not} mass eigenstate & ``special/single'' mode & external $N_R$ 
\tabularnewline
%
%\hline
%
Mass for $c_N > -1/2$ & $M_N^{ \rm UV } \times
\left( \frac{ \rm TeV }{ M_{ \rm Pl } }\right)^{ -2 \; c_N - 1 }$ 
& $M_N^{ \rm UV } \times 
\left( \frac{ M_N^{ \rm UV} }{ M_{ Pl } } \right)^{ \frac{1}{ -2 \; c_N } - 1 }$ 
& $M_N^{ \rm bare }
\left( \frac{ \mu }{ M_{ \rm Pl } } \right)^{ 5 - 2 \big[ {\cal O}_N \big] }$ 
\tabularnewline
Mass for $c_N < -1/2$ & $M_N^{ \rm UV }$ & $M_N^{ \rm UV }$ & $M_N^{ \rm bare }$ \tabularnewline
fraction of (net) $m_{ \nu }$ & $1$ & $\ll 1$ for $c_N > -1/2$ & $0$ \tabularnewline
& & $\gg 1$ (``cancels''  $\gg1$ below) for $c_N < -1/2$ & 
\tabularnewline
\hline
fraction of (net) $m_{ \nu }$ from & $0$ & $\ll 1$ 
for $c_N > -1/2$ & 
?!
%
%$\ll 1$ 
%
(for {\em both} $c_N <-1/2$ and $>-1/2$)
\tabularnewline
{\em sum} of intermediate modes & & $\gg 1$ for $c_N < -1/2$  & 
%
%(from Dirac 
%
%state/
%
%modes) & (from pseudo-Dirac {\em pair}s) & (from Dirac composites) 
%
\tabularnewline
\hline
\end{tabular}
} 
\caption{A comparison of the three bases used for studying this model. Note that the bulk mass for singlet field in the
5D model ($c_N$) is dual (in the CFT picure) to $\left(2-\big[{\cal O}_N \big] \right)$, where $\big[{\cal O}_N \big]$ is the
scaling dimension of the singlet operator in the CFT basis.
%
%$\big[{\cal O}_N \big]$. 
%
Whereas, the Majorana mass on the Planck brane in the 5D model ($M_N^{ \rm UV }$) corresponds to the bare mass for the external singlet ($M_N^{ \rm bare }$) in the
CFT interpretation.}
\label{summary_table}
\end{adjustwidth}
\end{table}

Here is the outline for the rest of this paper. We begin with a review of the above 5D model, setting-up our notation in section \ref{model}. 
%As mentioned above, our goal is to fill the above-mentioned gaps in the studies of this warped seesaw model, thereby providing a more complete picture. \sh{can be ``stronger'' than this ?}
%
In order to set the stage for our new analysis, it is necessary to first give a more extensive review of the various related results from earlier literature, namely, that of the KK basis calculation done earlier. We do this in section \ref{KK}.
% by giving a {\em qualitative} summary of the relevant features first (subsection \ref{KK_qual}), and then providing the actual formulae in subsection \ref{KK_quant}. 
We then move onto {\em our} findings. 

Our mass basis calculation of the SM neutrino mass is given in section \ref{massbasis}; this is a somewhat tedious procedure and so we begin (subsection \ref{mass_summary}) with a qualitative {\em summary} of the subsequent results, followed by setting-up the mass basis in subsection \ref{mass_setup}. The main results are summarized in subsection \ref{mass_result}. In table \ref{summary_table} we give a snapshot of the features in each of the three bases mentioned above (KK basis, mass basis CFT basis). Each entry will be clarified below. The full details of 5D calculation are relegated to Appendix~\ref{tedious}.

In section \ref{CFT} we scrutinize the 5D model from a 4D CFT perspective, starting (again) with a brief summary followed by more detailed subsections. We finally present our conclusions in section \ref{conclude}, where we also discuss some directions for future work.

%
%In appendix \ref{KKdiagmass}, we briefly discuss/outline how to go to the mass basis by diagonalization of mass matrix for the
%KK modes.
%

\section{The 5D Model}
\label{model}

We consider a slice of $AdS_5$ geometry described by the following metric:
\beq \label{eq:metric}
d s^2 = \left(\frac{R}{z}\right)^2\eta_{ab}~dx^{a} dx^{b},
\eeq
where $\eta_{ab} = \text{diag} (+, -, - ,- , -)$ and $x^a=(x^\mu,z)$, with $\mu=0,1,2,3$ and the fifth coordinate confined within the interval $R \leqslant z \leqslant R'$.~\footnote{As a reference it is useful to recall that much of the literature uses the equivalent line element $d s^2 = e^{-2ky} \eta_{\mu\nu} dx^{\mu} dx^{\nu} - dy^2$, with $0\leq y\leq \frac{1}{k}\ln(kR')$ related to ours by $z = \frac{e^{ky}}{k}$ and $k=1/R$.} At the boundary $z=R~(R')$ we locate a UV (IR) brane.
The SM fermions are in the bulk and, for simplicity, the SM Higgs boson is taken to be localized on the IR brane, although we think that the arguments presented here can be straightforwardly generalized, giving similar results, as long as the Higgs boson is peaked towards the IR brane. 

In order to be consistent with bounds from EW precision tests, we consider a minimally extended bulk gauge group $SU(2)_L \times SU(2)_R \times U(1)_{B-L}$ with $SU(2)_R \times U(1)_{B-L}$ spontaneously broken down to $U(1)_Y$ on the UV brane. Since detailed dynamics responsible for such a spontaneous breaking is not of central interest here, we will not discuss it for brevity. However, it is worth to mention that in this framework the SM singlet neutrino is charged under $SU(2)_R \times U(1)_{B-L}$. Since the Majorana mass term for the singlet breaks this gauge symmetry it can appear only on the UV brane. 

The quadratic action for SM singlet neutrino \footnote{For simplicity, we describe {\em one} generation, but our analysis can be easily extended to more.
%
%is rather straightforward. 
%
}
in the background (\ref{eq:metric}), including a UV-localized Majorana mass ($S_{\rm UV}$), is:
\ba \label{eq:action_1}
S &=& \int d^5 x \sqrt{g} \left\lbrace \frac{i}{2} \left( \bar{\Psi} e_a^M \gamma^a D_M \Psi - D_M \bar{\Psi} e_a^M \gamma^a \Psi \right) - m_D \bar{\Psi} \Psi  \right\rbrace +S_{\rm UV}\\\no
 &=&\int d^5x \left( \frac{R}{z} \right)^4 \left\lbrace -i \bar{\chi} \,\bar{\sigma}^{\mu} \partial_{\mu} \chi - i \psi \, \sigma^{\mu} \partial_{\mu} \bar{\psi} + \frac{1}{2} \left( \psi \overleftrightarrow{\partial_5} \chi - \bar{\chi} \overleftrightarrow{\partial_5} \bar{\psi} \right) + \frac{c_N}{z} \left( \psi \chi + \bar{\chi} \bar{\psi} \right) \right\rbrace+S_{\rm UV}.
\ea
In the first line the F\"{u}nfbein reads $e_M^a=(R/z)\delta_M^a$, $D_M=\partial_M+\omega_M$ with the spin connection given by $\omega_M = \left( \frac{\gamma_{\mu} \gamma_5}{4 z}, 0  \right)$. For the gamma matrices we use the conventions of~\cite{Csaki:2003sh}:
\beq \label{eq:gamma_matrix}
\gamma^{\mu} = \left(
\begin{matrix}
0 & \sigma^{\mu} \\ \bar{\sigma}^{\mu} & 0 
\end{matrix} \right)
\quad\quad \sigma^0 = -1, 
\quad \gamma^5 = \left(
\begin{matrix}
i 1 & 0 \\ 0 & -i 1
\end{matrix} \right).
\eeq
In the second line we explicitly wrote the action in terms of Weyl spinors:
$$
\Psi= \left( \begin{array}{cc} \chi_{\alpha} \\ \bar{\psi}^{\dot{\alpha}}\end{array} \right),
$$
and defined the real number $c_N\equiv m_DR$, and $\overleftrightarrow{\partial_5} \equiv \overrightarrow{\partial_5} - \overleftarrow{\partial_5}$.

The UV-localized Majorana mass term is defined as a quadratic term for $\psi$:
\beq \label{eq:action_majorana}
S_{\rm UV} = \int d^5x \left( \frac{R}{z} \right)^4 \frac{d}{2} \delta (z-R) \frac{R}{z} \psi \psi + {\rm hc},
\eeq
where $d\equiv M_N^{\rm UV}R$.

We also introduce a coupling between $\Psi$, a Higgs ${\cal H}$ localized on the IR-brane at $z=R'$, and the electroweak doublet 5D field $\Psi_L$:
\ba\label{eq:4D_Yukawa_eff_1}
\delta S &=& - \int d^4x \int dz \left(\frac{R}{z}\right)^4\delta (z-R')\lambda_5 {\cal H} \Psi_L \Psi 
\ea
where $\lambda_5$ is 5D Yukawa coupling with mass dimension -1. 
In our notation $c_{ N, L }$ denote the 5D mass parameters for RH (singlet) and LH (doublet) neutrinos (which, in turn, determine profiles for zero-modes in the extra dimension). We will follow convention that  $c_L = 1/2$ ($c_N=-1/2$) is constant profile for the LH (RH) zero mode, $c_L > 1/2$ ($c_N<-1/2$) being localized close to the Planck/TeV brane. Values $c_L \gtrsim 1/2$ are expected to explain the smallness of the charged lepton masses~\footnote{There might be some leeway here, due to the profile of RH charged lepton. In any case, formulae below can be easily generalized to $c_L < 1/2$ by replacing $\sim \left( \hbox{TeV} / M_{ \rm Pl } \right)^{ c_L - 1/2 }$
by $\sim \sqrt{ 1/2 - c_L }$.}

All dimensionful parameters are taken to be $O(1)$ in units of AdS curvature scale and in turn, the latter mass scale is set to be the 4D Planck mass scale. In the following, by ``TeV scale'', we tacitly mean the scale $1/R'$ which sets size of KK masses.

\section{SM neutrino mass using KK basis}
\label{KK}

In this section, we will first 
%
%briefly 
%
review previous results obtained using what we call the {KK basis} and present our new work
in the following section.
As outlined in the introduction, this KK basis is characterized by 
an {\em a-posteriori} consideration of the effects of the UV brane Majorana mass term on the modes (both zero and massive KK) which had been 
%
%at first 
%
obtained with{\em out} this UV brane mass term: 
essentially this ``addition'' generates Majorana mass terms for all these modes:
see, for example, reference \cite{Huber:2003sf}.\footnote{Note that in the literature, there are
usages of ``KK'' basis/modes with other meanings, for example, while dealing with charged fermions (i.e., no Majorana mass!), some authors denote by it the mass/physical 
basis/modes {\em before} taking into account EWSB (Higgs VEV), i.e., doublet and singlet modes are separate, 
whereas some others reserve it for the final, i.e., post-EWSB, 
physical/mass basis. Once again, our KK basis for {\em singlet} is the one with{\em out} taking into account {\em both} Majorana mass term on Planck brane {\em and} mass mixing with doublet leptons via EWSB.} 
To begin with, we provide a simple derivation -- using equations of motion (EOM) -- 
%for how exchange of $N$ would-be zero mode (only), i.e., not the KK modes, contributes to 
of the SM neutrino mass. The result that we are about to derive was already obtained and used in earlier works \cite{Huber:2003sf, Perez:2008ee}, but with different method. Rather than following the approach used in the literature we present a different derivation, that makes the relevant physics more transparent.

\begin{figure}
\begin{adjustwidth}{-0cm}{-0cm}
\centering
\mbox{\includegraphics[height=50mm]{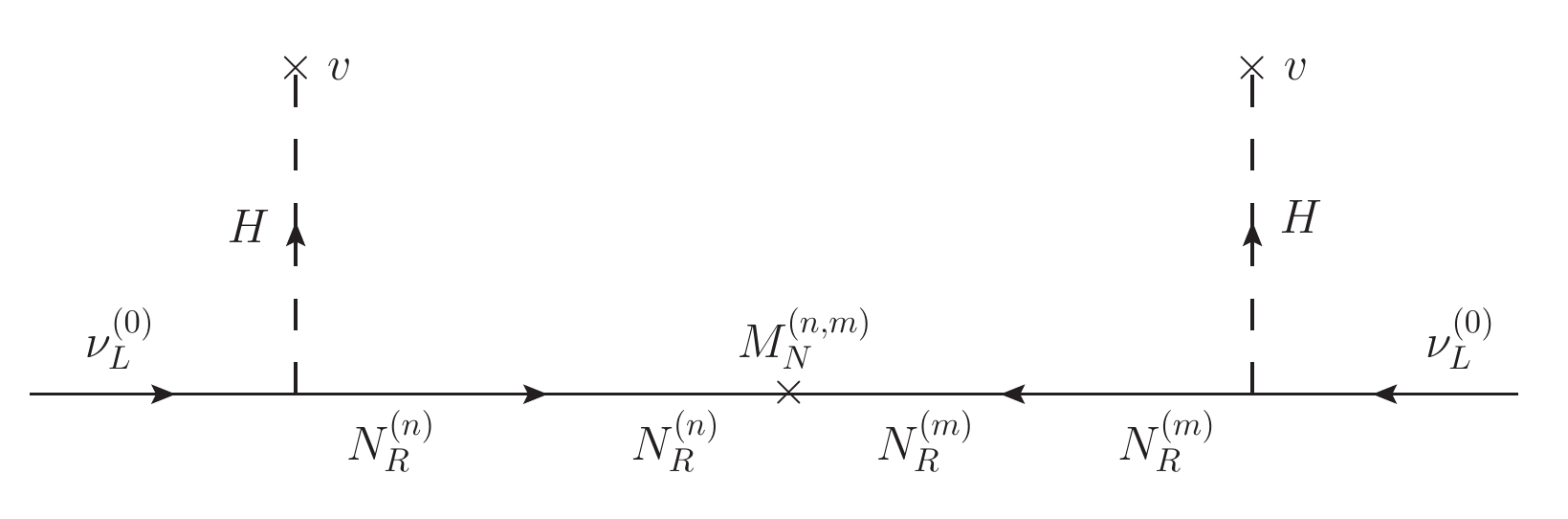}}
\caption{{The (vanishing) SM neutrino mass contribution from exchange of 
massive/KK modes in KK basis, where $M_N^{ (n,m) }$ $( n, \, m \neq 0$) 
denote Majorana mass terms.}
\label{nu_mass_KK}}
\end{adjustwidth}
\end{figure}

%\begin{figure}

%\vspace{-10cm}

%\hspace{-2.5cm}

%\includegraphics{nu_mass_KK}

%\vspace{-15cm}

%\caption{The (vanishing) SM neutrino mass contribution from exchange of 
%massive/KK modes in KK basis, where $M_N^{ (n,m) }$ $( n, \, m \neq 0$) 
%denote Majorana mass terms.}
%\label{nu_mass_KK}
%\end{figure}
%

We use 4-component Dirac spinors notation, with $N_R
%
%_R
%
^{ (0) }$ being singlet {\em chiral} 
%(i.e., right-handed only)
zero-mode
% [there is {\em no} $N_L^{ (0) }$]
, $N^{ (n \neq 0) }$ being singlet {\em non}-zero KK modes (Dirac i.e. have both L and R chiralities)
%, made of two distinct Weyl spinors) 
and 
$\nu^{ (0) }_L$ being (doublet) SM neutrino (left-handed only). We have the following mass terms 
\bea
{\cal L}_{ \rm mass } & = & \sum_{ n , m = 0, 1,2...} \frac{1}{2} M_N^{ (n, m) } \overline{ \Big[ N^{ (n) \; c } \Big]_L } N_R^{ (m) }  + \sum_{ n = 
1, 2...} m_n \overline{ N^{ (n) }_L } N_R^{ (n) } + \sum_{m=0, 1 ...} m_D^{ ( 0, m ) } \overline{ \nu^{ (0) }_L } N_R^{ (m) }  \nonumber \\
& & + \hbox{ h.c.}
\label{LKK}
\eea
where $m_D^{ ( 0, m ) }$ is the (effective) 
Dirac mass for the two {\em different} types of neutrino 
%(i.e., doublet and singlet) 
modes induced by the Higgs VEV. These EWSB-induced mass terms are given simply by 5D Yukawa coupling (along with
Higgs VEV) multiplied by product of profiles of LH (zero) and RH (zero or KK, labelled $m$) neutrino modes at the IR brane.
Similarly, 
$M_N^{ (n, m)}$ are Majorana mass terms between various singlet 
%(only)
modes, obtained by multiplying the
Majorana mass term on the UV brane by relevant profiles at the UV brane.
Finally, $m_n$ are the usual Dirac masses for the non-zero KK modes.
\footnote{In \cite{Huber:2003sf} the Dirac masses are denoted by $D_n$ (our $m_n$). The Majorana mass terms between singlet modes, which we denoted as $M_N^{ (n, m)}$, is denoted $A_{nm}$. Finally, the Dirac mass between LH zero mode and RH zero/KK modes, which we called $m_D^{ ( 0, m ) }$, is denoted $C_{0n}$ in \cite{Huber:2003sf}.}

We simply use 
equation of motion for $N_L^{ (n \neq 0) }$ which implies
$N_R^{ (n \neq 0)} = 0$, since only term in Lagrangian involving $N_L^{ (n) }$ is the KK mass with $N_R^{ (n) }$. 
Whereas, EOM for $N_R^{ (0) }$ sets itself to $\nu^{ (0) }_L m_D^{ ( 0,0 ) } / 
M_N^{ ( 0, 0 ) }$. Plugging these expressions for $N_R^{ (n) } \; (n= 0, 1...)$ back into the
Lagrangian we get
\bea
{\cal L} & \ni & - \frac{1}{2} \frac{ \Big[ m_D^ { ( 0, 0  ) } \Big]^2 }{ M_N^{ ( 0, 0 ) } } \overline{  \nu^{ (0) }_L } \Big[ \nu^{ (0) \; c } \Big]_R
\eea
Equivalently, we can represent the
use of EOM's by Feynman diagrams (or use Feynman diagrams as ``mnemonic'' for EOM's), see Fig.~\ref{nu_mass_KK}.
In this KK basis, it is the right chirality of the KK mode 
which couples to both Higgs VEV at one end and has
Majorana mass term on the other side. Thus, we have to pick the ``$p \! \! \! \! \! \not \; \; $'' piece of propagator, which does {\em not} contribute at leading order (again, despite the non-zero KK modes having Majorana mass terms)
This argument is {\em not} valid for $N_R^{ (0) }$, so the entire contribution comes from the would-be zero-mode.

The formula for the SM neutrino mass from the would-be zero mode exchange looks like the usual, type I seesaw, i.e., 
\bea
m_{ \nu } & \equiv & \frac{ m_D^{ \hbox{eff} \; 2 }}{ M^{ \rm eff}_N }
\label{master}
\eea
where $m^{ \rm eff}_D = m_D^ { ( 0, 0 ) }$ for the case of would-be zero mode, with 
\bea
m_D^{ ( 0, 0 ) } & \approx & 
\left\{ 
\begin{array}{ll}
a_{ > -1/2 } 
Y_5 \; v \left( \frac{ \rm TeV }{ M_{ \rm Pl } } \right)^{ c_L - \frac{1}{2} } 
& \hbox{for} \; c_N > -\frac{1}{2}  \\
a_{ < -1/2 } 
Y_5 \; v \left( \frac{ \rm TeV }{ M_{ \rm Pl } } \right)^{ c_L - \frac{1}{2} } \times
\left( \frac{ \rm TeV }{ M_{ \rm Pl } } \right)^{ -c_N - \frac{1}{2} } 
& \hbox{for} \; c_N < -\frac{1}{2}
\end{array}
\right.
\label{mD0}
\eea
where the superscript $(0,0)$ on $m_D$ indicates that this is the mass term between two zero modes, obtained by combining their profiles at the TeV brane 
(we assumed $c_L > 1/2$ for simplicity here).
Also, $Y_5 \equiv \lambda_5/R$ denotes the Yukawa coupling of brane-localized Higgs to bulk fermions in units of AdS curvature scale ($k$).
Here (and in what follows), we have kept track of {\em parametric} effects, i.e., relegating the $O(1)$ factors
to seperate formulae:
\bea
a_{ > -1/2 } & \approx & \sqrt{ \frac{(2c_N+1)(2c_L-1)}{2} } \\
a_{ < -1/2 }  & \approx & \sqrt{ \frac{(-2c_N-1)(2c_L-1)}{2} }
\eea
Similarly,  
the effective Majorana mass 
%
%here
%
in Eq.~(\ref{master}) is given by the Majorana mass {\em term} of the  would-be zero mode {\em with itself}, $M_N^{ \rm eff} = M_N^ { ( 0, 0 ) }$
\footnote{We emphasize that (see also next section) these KK basis modes are {\em not} the mass eigenstates; in order to make this point
explicit, we denote this mass term as above, instead of simply $M_N^ { ( 0 ) }$, which would give the impression that it is actually a
physical {\em mass}.}, with
\bea
M_N^ { ( 0, 0 ) } 
& \approx & 
M^{ \rm UV }_N \times \left\{ 
\begin{array}{ll}
b_{ > -1/2 } 
\left( \frac{ \rm TeV }{ M_{ \rm Pl } }\right)^{ 1 +  2 \; c_N } 
& \hbox{for} \; c_N > -\frac{1}{2} \\
b_{ < -1/2 }
& \hbox{for} \; c_N < -\frac{1}{2}
\end{array}.
\right.
\label{MN0}
\eea
namely, size of Majorana mass term on UV brane, denoted by $M^{ \rm UV }_N$, multiplied by (square of) the  profile of the would-be zero mode for the RH neutrino at the 
UV brane this time. Once again, $b$'s above are $O(1)$ factors, given by
\bea
b_{ > -1/2 } & \approx & \left(  2 c_N +1 \right) \\
b_{ < -1/2 }  & \approx & -\left(2 c_N +1 \right)
\eea
Plugging the singlet would-be zero mode Majorana mass from Eq.~(\ref{MN0}) and its Dirac mass with doublet zero mode from
Eq.~(\ref{mD0}) into the ``master'' formula in Eq.~(\ref{master}), we get (for {\em both} $c_N <$ and $> -1/2$) 
\bea
m_{ \nu } \approx \left( c_L - \frac{1}{2} \right) \frac{Y_5^2 v^2}{ M^{ UV }_N } \left( \frac{ \hbox{TeV} }{ M_{ \rm Pl } } \right)^{ 2 \left( c_L - c_N - 1 \right) }
\label{mnu5D}
\eea
As promised, deriving formula for the SM neutrino mass is a very straightforward task in KK basis!

It is remarkable that the strong dependence on $c_N$ is similar whether we consider $c_N < -1/2$ or $c_N > -1/2$. This requires more explanation. First of all, as can be seen from Eq(\ref{mD0}), for $c_N < -1/2$, the Dirac mass is exponentially suppressed by the fact that the profile of RH singlet would-be zero mode is peaked at UV brane and highly suppressed at IR brane. On the other hand, the Dirac mass for $c_N > -1/2$ does not show any strong sensitivity in $c_N$, which again comes from the fact that the profile at IR brane is unsuppressed and has very little $c_N$-dependence in this case. In the case of Majorana mass, however, the situation is \emph{interestingly} reversed (see Eq(\ref{MN0})). Namely, it is now $c_N > -1/2$ case that acquires exponential suppression and only a mild $c_N$-dependence for $c_N < -1./2$. After combining these two effects, one can now, at least intuitively, see that in both $c_N <$ and $>-1/2$ cases the SM neutrino mass gets strong $c_N$ dependence as explicitly shown in Eq(\ref{mnu5D}). What's really remarkable is that everything works out just right such that both cases reveal exactly the same $c_N$-dependence. In section \ref{CFT}, we will come back to this point and provide another way to understand it in a somewhat less coincidental manner. The above-mentioned results in KK basis are summarized in table \ref{summary_table}.

%although in the former case, it is due to the profile of the would-be zero mode of the RH neutrino on the {\em TeV} brane appearing in the  Dirac mass (numerator), with the Majorana mass, i.e., the profile on the UV brane (denominator) being rather insensitive to $c_N$.
%
%vs.~
%
%Whereas, for $c > -1/2$, it is the converse situation, i.e., sensitivity to $c_N$ derives from the Majorana mass, i.e., the profile of the would-be zero mode on the Planck brane instead (with that on the TeV brane, i.e., contributing to the Dirac mass, having much milder dependence on $c_N$).
%We will return to this rather curious point later on.

Before moving to a study of the mass basis, we stress that in type I high-scale seesaw models (including the 5D realization above) there {\em appears} to be a ``new hierarchy'' of mass scales. This is because the
(effective) seesaw scale needed is $\sim O \left( 10^{ 12} \right)$ GeV, i.e., $\sim 6$ orders of magnitude smaller than Planck scale\footnote{In other words, it is {\em not} enough to get small $m_{ \nu }$ -- which is accomplished by the basic seesaw mechanism for {\em any} high scale for singlet neutrino mass,
but we need to get its right size as well, which requires seesaw scale to be high, but not as much as Planck scale!}.
In order to achieve this in the 4D models, one is usually forced to introduce {\em new} dynamics for this purpose, often requiring its own explanations. This is what would also happen in our model if we took $M_N^{ \rm UV } \ll M_{ \rm Pl }$. Importantly, in warped 5D models there is an interesting alternative. In fact, the desired seesaw scale can be
obtained from Planckian-size $M_N^{ \rm UV }$ {\em naturally}, it suffices to choose $|c_N|$ a bit smaller than $1/2$ for $M_N^{ \rm eff}$ to be 
(much) smaller than the Planck scale. Specifically, in order to get the observed size of the SM neutrino masses, given that $c_L \sim 0.6$ is a ``natural'' choice\footnote{i.e., it can
account for charged lepton mass hierarchies {\em and} 
suppress flavor violation with{\em out} any significant structure 
%
%hierarchies 
%
in the 5D Yukawa couplings, in addition to being safer from EW precision tests than $c_L < 1/2$.} for reproducing charged lepton masses 
[i.e., $m_D^{ (0,0) }  \sim O( 10 \; \hbox{GeV})$]\footnote{Note that
this (i.e., neutrino) Dirac mass is only suppressed by {\em one} factor of doublet lepton profile, cf.~charged lepton mass
involving two such factors; that is why we can take $O( 10 \; \hbox{GeV})$ as
Dirac mass term for neutrino, instead of $\sim O ( \hbox{GeV} )$ for charged lepton, say, $\tau$, mass.}, we can 
choose $c_N \sim -0.3>-1/2$ so that for natural size of $M_N^{ \rm UV }$ [namely $\sim O(M_{ \rm Pl })$],
we get $M_N^{ \rm eff} \sim O \left( 10^{ 12 } \right)$ GeV, giving us $m_{ \nu } \sim O( 0.1 )$ eV as required.

\section{SM neutrino mass using mass basis}
\label{massbasis}

The reader must be warned that the KK basis is {\em not} even remotely close to the mass basis. Indeed, the Majorana mass term for low-lying (TeV-scale) KK modes can be much larger than KK (Dirac) mass itself:
\bea
M_N^{ (1,1) } & \sim & M^{ \rm UV }_N \times 
\left\{
\begin{array}{l}
\left( c_N + \frac{1}{2} \right)^2
\left( \frac{ \hbox{TeV} }{ M_{ \rm Pl } } \right)^{ - 2 \; c_N - 1 }, \; \hbox{for} \; c_N < -1/2
\\
\left( c_N + \frac{1}{2} \right)^2 \left( \frac{ \rm TeV }{ M_{ \rm Pl } }\right)^{ 2 \; c_N + 1 }, \; \hbox{for} \; c_N > -1/2
\end{array}
\right.
\label{Majorana11}
\eea
where we are interested in $ c_N \sim -1/2$ and $M_N^{ \rm UV } \lesssim M_{ \rm Pl }$ so that 
(typically) $M_N^{ (1,1) } \gg$ TeV.
This demonstrates that the Majorana mass terms can{\em not} 
really be treated as a ``perturbation'' (i.e., that it should be included from the beginning).

We therefore decide to analyze the warped seesaw model 
using {\em mass} basis {\em directly}, which is necessary for the study of direct production of singlet neutrino states in the early universe (relevant perhaps for leptogenesis) or  
at colliders. Namely, we 
include the effect of the Majorana mass on the Planck brane a {\em priori} such that all modes are (from the start) 
Majorana\footnote{Strictly speaking and as mentioned earlier, EWSB will actually further mix the singlet modes in this ``mass'' basis with doublet modes, 
but {\em that} 
effect can be genuinely treated as a perturbation, just like it is often done for charged SM fermions: we will neglect it -- at this stage -- for simplicity
and so continue to call it the mass basis, again for the singlet modes {\em by themselves}. Of course, these EWSB-induced mass terms between singlet and doublet zero-mode, i.e., the SM neutrino are crucial later, i.e., in generating mass for the SM neutrino.}. The two approaches must of course agree on the final result. Nonetheless, 
we will see that this change of basis has some ``surprises'' in store for us that will elucidate the nature of the seesaw mechanism itself! An intuitive understanding of our results immediately follows from the CFT interpretation 
in section \ref{CFT}.

%
%and appendix 
%\ref{KKdiagmass} for an approach for diagonalization starting from the more familiar KK basis in order to obtain the mass eigenstates.
%

\subsection{Summary}
\label{mass_summary}

We first give highlights of the mass basis analysis, before entering quantitative details in the next subsection.

It turns out that basically all modes (except one) are ``pseudo-Dirac'', i.e., form pairs with (roughly) the ``original'' Dirac-like mass, but with very {\em small} mass splitting within each pair, 
induced by the Majorana mass term on the UV
brane.
This spectrum comes with
a
%
%usual
%
regular spacing between these pairs, given by $\sim$ TeV (the usual KK scale): 
%in short,
in other words, each $\sim$ TeV interval (starting at $\sim$ TeV itself) 
in mass has 2 almost degenerate Majorana modes.
In addition to the mass spectrum, we need to know the {\em couplings} to Higgs (and doublet lepton) of these singlet modes;
%(which will result in EWSB-induced mass terms between these two types of neutrino modes);
they turn out to be {\em sizable}, given the localization of these mass eigenstates near TeV brane. %(where the Higgs boson also lives).
These two properties (which are {\em qualitatively} similar for both 
$c_N <$ and $> -1/2$) can then be combined as done above in the KK basis in order to get the SM neutrino mass.

We find that using the {\bf mass basis} 
%
%provides 
%
points to 
a strikingly different 
%
%picture of the
%
underlying mechanism 
%or structure'' (if you will) 
of the generation of SM neutrino mass, giving the same end result for the SM neutrino mass itself. 
%
%Once again, we only mention/list 
%
%summarize 
%
%the main points here (with details given 
%in the following subsections).
%
First of all, 
%
%there is a surprise in store about the type of seesaw itself 
%
%(again, repeating the point from the introduction):
%
%\ka{again, I am Ok with removing the following bullet point!}):
%
%\begin{itemize}
%
%\item
%
in the mass basis, the contribution of $\sim$ {\em TeV} mass singlet states 
%(based on the above-mentioned properties) 
to the SM neutrino mass 
%
%[via their -- (very) small -- Majorana mass splitting mentioned above]
%
is similar in size (for both 
$c_N <$ and $> -1/2$) to the final result.
%(which was indicated in the previous section, as obtained using the KK basis). 
%
Thus, even though it  ``started out'' trying to be type I, 
%
%this finding, in conjunction with the properties of these states mentioned above 
the {\em same} 5D model (again, in the mass basis) is 
reminiscent 
%(or has 
%
%features/
%
%characteristics
%
%component)
%
%ingredients
%
%of
%
%looks like/mimics 
%
the so-called
%
%dubbed
%
``inverse'' seesaw mechanism in the context of (purely) 4D models 
%
%mechanism 
%
\cite{Mohapatra:1986aw}. Namely, {\em both} this 5D model and the 4D models in \cite{Mohapatra:1986aw} 
(and follow-ups) are characterized by SM neutrino mass
originating from exchange of a singlet mode(s) with 
very {\em small} Majorana mass term combined with its couplings to Higgs {\em not} being small!
%
%\end{itemize}
%
In other words, the
mechanism for
%
%a direct probe of
%
%understanding of 
%
the generation of SM neutrino mass  might be ``closer at hand'' than had been anticipated in the KK basis: for example, 
\begin{itemize}

\item
the {\em TeV} mass singlet states, whose exchange generates the SM neutrino mass, can potentially be probed 
at the LHC (or future colliders).

\end{itemize}
%{\color{green} Remove? (repeated below) Of course, one might have contemplated collider signals for these singlet neutrino states in any case (i.e., no matter what basis we use), given that their mass is $\sim$ {\em TeV}, i.e., within reach of the LHC/future colliders. However, in KK basis, first of all, such a study might not have been precise since it is not mass basis. In addition and perhaps more importantly, under the ``illusion'' that it is type I in this basis, one would {\em not} have thought that these modes provide a {\em direct} link to the generation of the SM neutrino mass, since that job would be (completely) relegated to the much heavier would-be zero mode instead (which is of course well beyond LHC/any foreseeable future collider reach) !}
%
Furthermore, 
\begin{itemize}

\item
for leptogenesis, the focus might now be on the  
%
%can occur or using 
%
decay of these TeV singlet states, which does {\em not} require the universe to be reheated
to temperatures (much) above a TeV, thus avoiding the issue of the (too slow) phase transition mentioned earlier.

\end{itemize}
Overall, we thus see that the mass basis picture leads to a dramatic {\em shift} in the expected phenomenology. Indeed, from the KK basis one might erroneously be drawn to conclude that the physics which generates the SM Majorana neutrino mass can{\em not} be probed {\em directly} at the LHC (or foreseeable colliders), and that leptogenesis would require the universe to be reheated to
temperatures (much) above a TeV, which might pose a problem in these scenarios.~\footnote{It is known \cite{Creminelli:2001th} that the transition from such a high-temperature phase (i.e., $\gg$ TeV) to the usual warped model below temperature of $\sim$ TeV might proceed {\em too} slowly, which might then become 
a bottleneck in implementing a standard 
leptogenesis scenario.
}
Our results show that none of this is true.

%
%as is the case
%for the usual (4D) inverted seesaw models
%
%than/vs.~for
%
%(cf.~high-scale type I)...but more natural here...
%

%
%essentially due to 
%small profile of TeV mass modes at the Planck brane
%
%\end{itemize}
%

Note that reference \cite{Fong:2011xh} actually added an extra (i.e., beyond the $N$ discussed above) singlet in the bulk to this model 
in order to implement inverse seesaw in 5D (which is the way it is done in usual, 4D models), but our claim here is that there is no ``need'' to do so. \footnote{In more detail,
in 4D inverse seesaw model, we consider {\em two} Weyl spinor singlets, which form a pseudo-Dirac state.
Reference \cite{Fong:2011xh} attempted to mimic this in the 5D model by incorporating two (chiral) zero-modes,
i.e., one from each of two (singlet) bulk fields. 
However, we see that such a ``proliferation'' of bulk singlets is actually 
not necessary since a {\em single} bulk field does have {\em two} chiralities at the {\em non}-zero mode level: we find 
that these form the required pseudo-Dirac state.}

Next, we mention finer points 
%(which were {\em not} listed in introduction) 
about the mass basis 
%(vs.~KK basis) 
analysis. For example, consider the ``fate'' (in the mass basis) of the would-be zero mode of the KK basis.
We can show that 
there is indeed one mode which is {\em un}paired:
% in the sense outlined
%
%mentioned 
%
%above, i.e., 
it seems to not conform to 
%
%amounts to an exception to 
%
the ``one {\em pair}-per-TeV bin'' rule. Hence, it is termed a ``special'' mode,
with what one might therefore call a ``purely'' Majorana mass. 
%
%Thus 
%
It is somewhat tempting to 
%
%somehow 
%
``identify'' it with the would-be zero mode of the KK basis discussed earlier. %
However, we find that this ``mapping'' is not quite accurate.
After a careful calculation, we discover that 
\begin{itemize}

\item

(i) for $c_N > -1/2$, the special mode in the mass basis is {\em not} at the would-be zero mode mass, but instead is 
parametrically {\em higher} (while still being smaller than the Majorana mass term on the UV brane), with a coupling to
the Higgs which is similar to would-be zero-mode however. 
Thus, its contribution to the SM neutrino mass 
%(which is of course of type I seesaw-type, cf.~inverse from paired modes)
 is {\em negligible}. Similarly, we can show 
that the effect of the (much) heavier than $\sim$ TeV {\em paired} modes is small, i.e., sum over
these mass eigenstates from bottom-up is convergent.
Hence, we can indeed say that the SM neutrino mass is 
{\em dominantly} of inverse seesaw nature, i.e., it basically 
 %
% does indeed 
 %
arises from exchange of $\sim$ TeV mass eigenstates mentioned above\footnote{Again, 
it {\em is} more than one pair of modes which contribute here, i.e., involving more like a ``tower'', albeit rapidly 
convergent, of inverse seesaws, but this is a minor variation with respect to the usual 4D model of this type.}.

(ii) $c_N < -1/2$: the special mode {\em is} in fact (roughly) at the would-be zero-mode mass. Nevertheless its coupling to Higgs is 
actually {\em un}suppressed, 
%(cf. that of would-be zero mode in the KK basis discussed above), 
giving
{\em too} large a contribution 
%(from this mode by itself) 
to the SM neutrino mass.
% (again, of type I seesaw).
However, we show that it 
is similar in size to the effect of the other, i.e., {\em higher} than $\sim$ TeV, ``special-paired''. We therefore {\em conjecture} 
%(as of now!) 
that these two contributions (again, each of them is {\em too} large) cancels against one another, leaving
behind that of the $\sim$ TeV modes mentioned above (which on its own is the ``correct'' size);
in this sense, we have sort of a 
``hybrid'' of inverse and type I seesaws here.

\end{itemize}

Finally, as far as the curious feature 
%(``puzzle'' if you will) 
about dependence on $c_N$ of the final SM neutrino mass
%in Eq.~(\ref{puzzle}) 
is concerned, 
we can boil it down to
\begin{itemize}

\item

the  dependence on $c_N$ of the Majorana mass splitting between the two ($\sim$ TeV) mass eigenstates in each pair being similar for $c_N > -1/2$ {\em and} $< -1/2$
(as mentioned above, this splitting is essentially what generates the 
%
%dominant or 
%
bottomline 
%
%contribution to the 
%
SM neutrino mass for {\em both} ranges of $c_N$).

\end{itemize}
The picture arising from our mass basis calculation is summarized in table 
\ref{summary_table}.

%%%%%%%%%%%%%
% technical details
%%%%%%%%%%%%%

\subsection{Setting-up the calculation}
\label{mass_setup}

We now show derivation of the above claims.
%/get ``dirty''!
%
Once again, in this approach, we take into account the Majorana mass term on the UV brane from the get-go so that {\em all} singlet modes are strictly speaking {\em Majorana}.
The calculation is rather straightforward, even if tedious: see Appendix~\ref{tedious} for details.
It turns out that these Majorana
mass modes can be divided into two types: light modes and special modes. The low-lying (TeV-mass) modes come {\em in pairs}
of pseudo-Dirac particles (a Weyl spinor with mass $m$ and another of mass $\sim-m$) and similar couplings to the SM Higgs and SM doublet 
neutrino. We will denote the two modes within each pair (and values of their masses and couplings) by the 
subscripts $\pm$, respectively.
Of course, we have an infinite tower of such modes, counted by $n = 1,2,\cdots$, so each $n$ actually stands for two, ``$\pm$'', modes.
In addition, 
at a mass scale much larger than $\sim$ TeV (essentially dictated by Majorana mass term on UV brane,
but appropriately modulated by profiles), we find an {\em un}paired/single
mode, which we dub ``special''.

\begin{figure}
\begin{adjustwidth}{-0cm}{-0cm}
\centering
\mbox{\includegraphics[height=50mm]{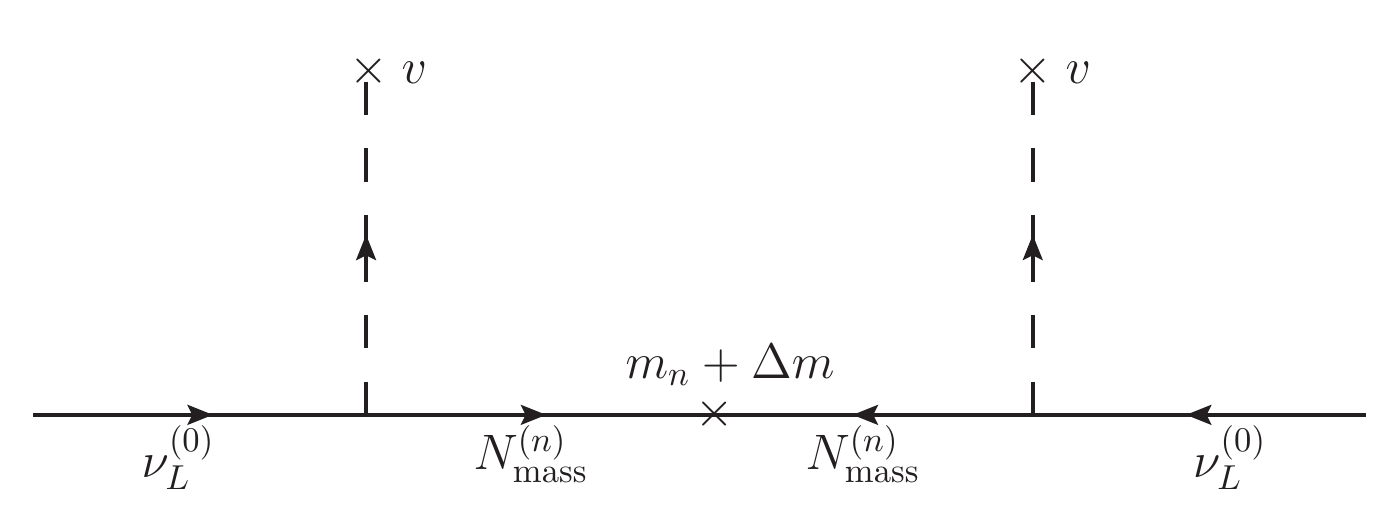}}
\caption{The SM neutrino mass from exchange of one singlet mode in mass 
basis, labelled $N_{ \rm mass }^{ (n) }$ and of mass 
$\left( m_n + \Delta m \right)$.}
\label{nu_mass}
\end{adjustwidth}
\end{figure}

The
single/special, Majorana mode (mass $M^{ \rm special }_N$, coupling $y^{\rm special}$ with Higgs and doublet neutrino zero-mode) gives the usual type I seesaw contribution to the SM 
%(i.e., doublet zero-mode) 
neutrino mass
\bea
m_{ \nu }^{ \rm special } & = & \frac{ \left( v \; y^{\rm special} \right)^2 }{ M_N^{ \rm special } } \nonumber
\eea
as in Fig.~\ref{nu_mass}, 
where $v \;y^{\rm special}$ is the Dirac mass 
with doublet neutrino zero-mode as usual.

{\em Each} mode of a pair of Majorana modes (mass $m_{ n \; \pm }$, {\em magnitude} of 
coupling $y_{n \; \pm }$) gives a contribution to the SM neutrino mass which is 
similar to the above. However, 
given the near-degeneracy with{\em in} each pair, it is convenient to consider their {\em combined} effect:
%
%contribution to the SM neutrino mass:
%
\bea\label{mass pair}
m_{ \nu }^{ \rm pair } & = & v^2 \left( \frac{ y_{n\;+}^2 }{ m_{ n \; + } } - \frac{ y^2_{n \; -}}{ m_{ n \; - } } \right) \nonumber \\
& \approx & \frac{y^2_nv^2}{m_n} \left(2\frac{\Delta y}{y_n}-\frac{\Delta m}{m_n} \right)
\eea 
again, as in Fig.~\ref{nu_mass}.
Here $\Delta y=y_{n\;+}-y_{n\;-}$ and $\Delta m=m_{n\;+}-m_{n\;-}$.

The procedure then is to determine the masses and couplings from a detailed 5D calculation, plug these into
above formulae, and finally sum over the pairs of Majorana modes.

\subsection{Results}
\label{mass_result}

In this section,
we will simply summarize the results of the above outlined procedure, referring the reader to the appendix \ref{tedious}
for the actual calculation.
As already mentioned in the summary above,
%
%introduction, 
%
each of the two cases $c_N >$ and $< -1/2$ has to be treated on its own.

\vspace{0.1in}

\noindent {\Large (i) $c_N > -1/2$} 

\vspace{0.1in}

We begin with the case of $ c_N > -1/2$, which is the phenomenologically {\em viable} option, i.e., can give the 
%
%observed 
%
known
size of the SM neutrino masses with natural choices of the bulk parameters. 
%(as indicated at end of previous section).

\vspace{0.1in}

 \noindent {\bf The {\em special} mode}  

\vspace{0.1in}

The 
%
%crucial point 
%
first surprising element 
is that the {\em mass} of special mode
%[obtained along the above lines: 
[for a derivation, see appendix \ref{sec:app_masses}\footnote{Following \cite{Huber:2003sf}, $M_N^{ \rm UV }$ in units of $M_{ \rm Pl  }$ is denoted by $d$ in
appendix \ref{tedious} also.}]  
%
%{\em not} at the value 
%
is 
{\em parametrically different} than 
the Majorana mass of the
would-be zero mode in the KK basis:
% that too, {\em parametrically} so:
%
%but is instead higher (that too, parametrically), 
%
namely, we find that 
\bea
M^{ \rm special }_N & \approx &
f_{ > -1/2 }
M^{ \rm UV }_N \times \left( \frac{ M_N^{ \rm UV} }{ M_{ Pl } } \right)^{ - \frac{1}{ 2 \; c_N } - 1 } 
\label{MNspecial1}
\eea
%
%
% which can be thought of as analog of...
%
with the $O(1)$ factor given by 
\bea
f_{ > -1/2 } & \approx & 2 \left( \frac{-\pi \tan (c_N \pi)}{\Gamma^2 (-c_N + 1/2)} \right)^{\frac{1}{2c_N}}
\eea
i.e., it is smaller than the input of $M_N^{ UV }$ (given that $c_N > -1/2$, the exponent is {\em positive} and we assume
$M_N^{ UV } \lesssim M_{ \rm Pl }$ here), but it is {\em larger} than the would-be zero mode mass in 1st line of Eq.~(\ref{MN0}).
On the other hand, the {\em coupling} of special mode to the SM Higgs
%
%(i.e., the Dirac mass term with the SM doublet neutrino) 
%
{\em is} (roughly) {\em similar} to that of the 
would-be zero-mode (apart from the 
%
%possible 
%
absence of the $\sqrt{ 1/2 + c_N}$ factor [which anyway
is $\sim O(1)$]), i.e., the EWSB-induced Dirac mass with the SM doublet neutrino, $m_D^{ \rm eff }$, is approximately\footnote{The reason for this similarity is, in turn, that of the profiles, i.e., they are both leaning towards the IR brane. Although it might not be needed (given the expectation 
based on these profiles), for an actual derivation of this coupling, 
see appendix \ref{sec:app_couplings}.}:
%
%see 1st line of Eq.~(\ref{mD0})
%
\bea
m_D^{ \rm eff, special } \sim m_D^{ (0,0) } \; [\hbox{where} \; m_D^{ (0, 0) } \; \hbox{is 1st line of Eq.~(\ref{mD0})}].
\eea 
Thus it is 
%
%easy to see 
%
clear that special mode's contribution to SM neutrino mass is {\em too} small to reproduce Eq.~(\ref{mnu5D}).

\vspace{0.1in}

\noindent {\bf {\em Low}-lying modes} 

\vspace{0.1in}

It is the TeV-mass physical modes 
%(which again are ``almost'' Dirac) 
which shoulder the responsibility of 
generating the SM neutrino mass. 
Their Yukawa coupling to the Higgs and the SM lepton doublet is suppressed only by the latter's profile at the TeV brane, given that these singlet profiles are peaked near the TeV brane, %(just like for the special mode or the would-be zero mode or for that matter for typical KK modes), 
i.e.,
$m_D^{ \rm eff }$ is again similar to $m_D^{ ( 0, 0 ) }$ in 1st line of Eq.~(\ref{mD0}).
Naively, one might then expect 
%
%too 
%
a {\em too} 
large SM neutrino mass from exchange of these
%
%TeV-mass 
%
modes, given the $\sim$ TeV mass for these modes. However, the crucial point is that  
the fraction of 
%
%purely
%
(primordially) ``Majorana natured''-mass 
%(i.e., splitting in each pair originating from Majorana mass term on the UV/Planck brane, mentioned above) 
is {\em naturally} very {\em small}.
%, cf.~in {\em KK} basis, the Majorana mass term is (much) larger than $\sim$ TeV [see 
%Eq.~(\ref{Majorana11}].
%
%and appendix \ref{KKdiagmass}}: 
%
From the explicit 5D mass basis calculation we find that the mass and coupling splitting are given by (see appendix \ref{sec:app_masses})
\bea
\frac{ \Delta m }{ m_n } & \approx & h_{>-1/2} \frac{ \hbox{TeV} }{ m_n } \frac{1} { M^{ \rm UV }_N / M_{ Pl } }  
\left( \frac{ m_n } { M_{ \rm Pl } } \right)^{ -2 \; c_N } \; \hbox{{\em irr}espective of} \; c_N 
\nonumber  \\
& \approx & 
h_{>-1/2} \frac{1} { M^{ \rm UV }_N / M_{ Pl } }   \left( \hbox{TeV} / M_{ Pl } \right)^{ -2 \; c_N }, \; \hbox{for} \; m_n \sim \hbox{TeV}
% 
%\hbox{for} \; c_N < \frac{1}{2} 
%
\label{5Dmass_split}\\
\frac{\Delta y}{y_n} &=& - c_N \frac{\Delta m}{m_n}
\label{5Dcoupling_split}
\eea
%
%only valid for low-lying/TeV-mass modes...
%
%can show higher modes effect is sub-dominant...
%
where the leading order mass $m_n$ and coupling $y_n$ are given by
\bea
m_n &\approx& \left( n + \frac{1}{2} (1-c_N) \right) \pi \;\;  \left({\rm TeV} \right) \\
y_n &\approx& Y_5 
%
%\sqrt{\frac{2c_L-1}{1-(M_{\rm Pl}/{\rm TeV})^{1-2c_L}}} 
%
\sqrt{ 2c_L-1 }
\left( \frac{ \rm TeV }{ M_{\rm Pl} } \right)^{ c_L - 1/2 }.
\label{mnyn}
\eea
(assuming $c_L > 1/2$ as before).
The $\mathcal{O}(1)$ factor $h_{>-1/2}$ is given by
\bea
h_{>-1/2} \approx \frac{4^{c_N} \pi}{\Gamma^2 (-c_N + 1/2)}.
\eea
As is discussed in detail in section \ref{sec:app_masses}, this $\mathcal{O}(1)$ factor is valid for any low-lying modes with not so small $n$ and more precise expression that holds even for the first few modes can be found there.
   
Notice that the mass (and similarly coupling) splitting is clearly $\ll 1$, as long as $c_N < 0$ and $M_N^{ \rm UV } \lesssim M_{ \rm Pl }$, i.e., for a (very) 
{\em wide} range of parameter space.
(We would like to again emphasize here that the above estimate for Majorana mass splitting holds both for $c_N >$ and $< -1/2$.)
Equivalently, we can treat the small Majorana splitting ($\Delta m$) as a ``mass insertion'' in getting to the above result. It should be clear from Eq.~(\ref{5Dmass_split}) and Eq.~(\ref{5Dcoupling_split}) that the contribution from mass splitting is similar in size to that due to coupling splitting. 
%\footnote{In principle, we also have to take into account the fact that the Yukawa couplings -- and hence the Dirac mass terms with the SM doublet neutrino -- also have a (small) shift (induced by the input Majorana mass term) of {\em opposite} sign for these two modes in the pair, which will result in give a contribution to the SM neutrino mass even if masses were {\em not} split. However, in practice, we can show that  this effect of couplings is of similar size to that due to {\em mass} splitting discussed above: see appendix \ref{sec:app_couplings} and \ref{sec:app_SM neutrino mass for c_N > -1/2}. So, for simplicity, we consider only the effect of mass splitting in these estimates; of course, our exact results include both effects.}

%
%\begin{figure}

%\vspace{-10cm}

%\hspace{-2.5cm}

%\includegraphics{nu_mass}

%\vspace{-15cm}

%\caption{The SM neutrino mass from exchange of one singlet mode in mass 
%basis, labelled $N_{ \rm mass }^{ (n) }$ and of mass 
%$\left( m_n + \Delta M \right)$.}
%\label{nu_mass}
%\end{figure}
%

Summing over such modes, 
%
%(and ignoring here factors of 2), 
%
we find that SM neutrino mass formula becomes
\bea
m_{ \nu } \approx & h_{>-1/2} (2 c_N+1) \sum_n \frac{ \hbox{TeV} } { M^{ \rm UV }_N / M_{ Pl } } \frac{ (y_n v)^2 }{ m^2_n } \left( \frac{ m_n } { M_{ \rm Pl } } \right)^{ -2 \; c_N}
\label{mnusum}
\eea
where we used Eq.~(\ref{5Dmass_split}) and Eq.~(\ref{5Dcoupling_split}).
Approximating $m_n $ by $\sim n$ TeV, we can see that this sum goes as $\sim \left( n_{ \rm max }^{ 2 \; c_N - 1 } - 1 \right)$, where $n_{ \rm max}$ ($\gg 1$) denotes a {\em naive} 
cut-off on the sum approaching from $n = 1$. Thus this sum is convergent 
for $c_N > -1/2$, which implies that it is dominated by the {\em lightest}, i.e., $\sim$ TeV mass modes
(this argument is valid only for $c_N > -1/2$). This is one of our main results. As far as the quantitative aspect is concerned, 
as indicated earlier, the expressions for masses and couplings given above are a very good approximation for low-lying modes with not so small $n$. However, since, as we just learnt, the contribution from the first few modes is significant, a more careful treatment is needed to get a more reliable final result. We do this carefully in the appendix \ref{tedious}, and as can be found in section \ref{sec:app_SM neutrino mass for c_N > -1/2}, the final answer for SM neutrino mass by performing numerical sum with improved $\mathcal{O}(1)$ factor shows excellent
%
%a very good 
%
agreement with the result obtained in the KK basis.

Having established the above quantitative result,
%
%agreement, 
%
we
now turn our attention to its 
qualitative features.
%
%aspect. 
%
For this purpose, it is clear that we can simply focus on the contribution from the lightest TeV mode. By setting $m_n \sim$ TeV in Eq.~(\ref{mnusum}) and noticing that the Dirac mass $y_n v$ is approximately 
$
%y_n v \sim 
%
m_D^{ (0,0) }$ [compare 2nd line of 
Eq.~(\ref{mnyn}) with Eq.~(\ref{mD0})], we get for $c_N > -1/2$
\bea
m_{ \nu } & \sim & \; \frac{1} { M^{ \rm UV }_N / M_{ Pl } } 
\frac{ \left[ m_D^{ (0,0)} \right]^2  }{ \hbox{TeV } }  \left( \hbox{TeV} / M_{ Pl } \right)^{ -2 \; c_N }.
%
%\label{}
%
\eea
Clearly it has the same form as Eq.~(\ref{master}), where the ``effective'' Majorana mass in this case 
%
%being given 
%
can be {\em defined} by 
\bea
M^{ \rm eff }_N & \sim & M^{ \rm UV }_N \left( \hbox{TeV} / M_{ Pl } \right)^{ 1 + 2 \; c_N  } 
%
%\nonumber \\
%
%& \hbox{equivalently} & \hbox{TeV} / \frac{ \Delta M^{ \rm Majorana }_N }{ \hbox{TeV} } 
%
%\label{}
%
\eea
which is {\em identical} to the would-be zero mode mass in the KK basis [see 1st line of Eq.~(\ref{MN0})]. 
Thus, it is easy to see that we reproduce the KK basis result already at this estimate-level.
However, it is important to realize that there is no ``special'' physics at $M_N^{ \rm eff }$ 
%itself 
in the {\em mass} basis, this scale is just an ``illusion".
%
%mirage''.
%

\vspace{0.1in}

\noindent {\bf Modes {\em near} special mode} 

\vspace{0.1in}

%
%A dedicated analysis shows that the mass splitting for these modes is $\sim$ TeV. Hence it is easy to see that their contribution to the %SM neutrino mass is negligible by the fact that the mass (not mass splitting) of these modes are much larger than that of would-be zero %mode in KK basis, a needed size for getting right answer. By explicit calculation we checked that the contribution due to coupling %splitting is similar in size to that from mass splitting, and hence is also negligible. 
%Such a result was perhaps expected, based on the sum over low-lying modes being convergent, combined with the 
%special mode (by itself, i.e., unpaired) giving too small an effect.
%
Based on the sum over low-lying modes being convergent, combined with the 
special mode (by itself, i.e., unpaired) giving too small an effect, we can anticipate 
%
%estimate 
%
that the modes near special mode will have a very small contribution 
%
%negligibly 
%
to the SM neutrino mass.
Indeed a dedicated analysis of the mass {\em and} coupling splittings of these modes confirms this expectation.
Similarly, we can estimate that the
modes {\em much} above the special one also contribute negligibly.

\vspace{0.1in}

\noindent  {\Large (ii) $c_N < -1/2$}

\vspace{0.1in}

Finally, for the sake of completeness we also briefly comment on the case $ c_N < -1/2$, even though does not give the observed size of neutrino masses for natural values of the bulk parameters. 

\vspace{0.1in}

\noindent {\bf {\em Special} mode} 

\vspace{0.1in}

Here, a similar analysis [for a derivation, 
see appendix \ref{sec:app_masses}] shows that the special mode (in the mass basis) {\em is} indeed at the mass of the would-be zero-mode:
\bea
M_N^{ \rm special } & \approx f_{<-1/2} & M_N^{ \rm UV } \; \hbox{for} \; c_N < -\frac{1}{2} 
\label{MNspecial2}
\eea
with the $\mathcal{O}(1)$ factor given by
\bea
f_{<-1/2} \approx -\left( 2 c_N + 1 \right).
\eea
but there is more to it than meets the eye! Namely, it is not just the mass, but also the {\em coupling} to the Higgs 
%(i.e., Dirac mass terms with the SM doublet neutrino) which 
%
%enters 
% 
is a player in this game of
the
generation of SM neutrino mass. It turns out that the ``analogy'' between the special mode of mass basis and the would-be zero mode of 
KK basis, based on similarity in their masses, does {\em not} extend to their coupling to the Higgs: from the {\em detailed} 5D calculation (see appendix \ref{sec:app_couplings}), we find that the coupling of the special mode 
to the Higgs is
{\em not} suppressed by the factor of the would-be zero mode profile at the TeV brane 
%(cf.~the coupling of the would-be zero-mode discussed in the KK basis)
simply because the special mode is peaked near the {\em TeV} brane (instead of near Planck brane for the would-be zero mode).
So, this is a rather
{\em un}expected result:
see section \ref{CFT} for some ``understanding'' of it in the CFT basis. 
Thus, we have 
\bea
m_D^{ \rm special, single } & \sim & v \left( \frac{ \rm TeV }{ M_{ \rm Pl } } \right)^{ c_L - \frac{1}{2} } \nonumber \\
& \sim & \left( \frac{ M_{ \rm Pl } }{ \rm TeV } \right)^{ -c_N - \frac{1}{2} } m_D^{ (0, 0) } \; (\hbox{2nd factor is second line of Eq.~(\ref{mD0})}) \nonumber \\
& \gg & m_D^{ (0, 0) } 
\label{mDspecial2}
\eea
(where we have labelled it ``single'' -- in addition to special -- since it is after all an {\em unpaired} mode: further reasons will be made clear later).
In other words, it is actually similar to the Dirac mass term (with the SM doublet neutrino) of the would-be zero mode in the KK basis 
%(or the special mass eigenstate) 
for the {\em other} value of $c_N (> -1/2)$ 
[see {\em 1st} line of Eq.~(\ref{mD0}), even though we have $c_N < -1/2$ in this case]. Equivalently, it is (roughly) same as the coupling of 
the {\em non}-special or KK 
%
%{\em TeV}-mass 
%
modes, {\em ir}respective of $c_N$: again, the point is that {\em all} these modes are peaked near the TeV brane.
Substituting Eqs.~(\ref{MNspecial2}) and (\ref{mDspecial2}) as the effective masses into Eq.~(\ref{master}), we see that 
\bea
m_{ \nu }^{ \rm special, single } & \sim & \frac{v^2}{ M^{ UV }_N } \left( \frac{ \hbox{TeV} }{ M_{ \rm Pl } } \right)^{ 2 \left( c_L - \frac{1}{2}  \right) } \nonumber \\
& \sim & m_{ \nu } \; [\hbox{of Eq.~(\ref{mnu5D})}] \times \left(  \frac{ M_{ \rm Pl } }{ \rm TeV } \right)^{-2 \; c_N - 1 } 
\label{mnu_special_single}
\eea
i.e.,
the contribution of the special mode by itself is too {\em large} compared to the KK basis result of Eq.~(\ref{mnu5D}).

Nonetheless, there is no reason to ``worry'' here, since only after summing {\em all} mass eigenstates would the
result for the SM neutrino mass agree with that obtained using the KK basis.
So, we now proceed to considering the contribution of the other modes carefully.

\vspace{0.1in}

\noindent {\bf {\em Low}-lying modes} 

\vspace{0.1in}

%
%On the other hand, 
%
Let us start with the low-lying modes, i.e., much below the special (single) one.
We can show that the Majorana mass (and similarly coupling) splitting for these non-special modes 
-- for the case $c_N < -1/2$ being considered here -- is also given by Eq.~(\ref{5Dmass_split}) that we used for $c_N > -1/2$ earlier (see appendix \ref{sec:app_masses} and \ref{sec:app_couplings}).
Also, the Dirac mass with the SM doublet neutrino for these modes is similar to that of the special mode in
Eq.~(\ref{mDspecial2}): equivalently, to that for the low-lying modes for the case $c_N >-1/2$ (again, this is expected based on all these profiles being peaked near the TeV brane).
Thus, we see that the {\em lowest}  TeV-scale modes ({\em no} sum yet!)
give a contribution to the SM neutrino mass that is 
similar in form to that discussed above for $c_N > -1/2$.
In other words, it is clear that, even for $c_N < -1/2$, the 
first few mass eigenstates (by themselves) contribute to the SM neutrino mass at order unity.

However, {\em un}like for $c_N > -1/2$ that we studied earlier, for the case of $c_N < -1/2$, 
as we include more and more low-lying modes, the {\em sum} seems to actually ``diverge'' from this {\em bottom-up}
viewpoint: this is easy to see from the 2nd line of Eq.~(\ref{mnusum}), where sum is $\sim 
\left( n_{ \rm max }^{ -2 \; c_N - 1 } - 1 \right) \sim n_{ \rm max }^{- 2 \; c_N - 1 }$ for the case of $c_N <-1/2$.
Obviously, these modes then also give {\em too} large contribution to the SM neutrino mass:
\bea
m_{ \nu }^{ \rm non-special } & \sim & n_{ \rm max }^{-2 \; c_N - 1 } \times m_{ \nu } \; [\hbox{of Eq.~(\ref{mnu5D})}]
\eea

We can thus naturally hope that the above sum might
%mostly, i.e., 
(up to the contribution of lightest modes) cancel against the special (single) mode contribution [Eq.~(\ref{mnu_special_single})]
-- {\em both} being overly large. 
In order to check this possibility, let us estimate the above 
sum of modes 
%(again, below special mode) 
by cutting it off at (roughly) mass of the special mode itself, i.e.,
set $n_{ \rm max } \sim M^{ \rm UV }_N / \hbox{TeV}$: this might be a reasonable way to proceed, since we do expect properties of modes to change as we make the transition across the special mode mass.
This assumption gives
\bea
m_{ \nu }^{ \rm non-special } & \sim & \left( \frac{ M^{ \rm UV }_N }{ \rm TeV } \right)^{ -2 \; c_N - 1 } \times m_{ \nu } \; [\hbox{of Eq.~(\ref{mnu5D})}]  \nonumber \\
& \sim & m_{ \nu }^{ \rm special, single } \times \left( \frac{ M^{ \rm UV }_N }{ M_{ \rm Pl } } \right)^{- 2 \; c_N - 1 }
\label{mnu_non_special}
\eea
where in 2nd line above, we have used Eq.~(\ref{mnu_special_single}). So, even though 
%(and, as expected) 
the collective effect of the light modes is much {\em larger} 
than the ``right'' answer, $m_\nu$, 
%(that we know from KK basis!),
%it still ``falls short'' (that too, {\em parametrically} so)
it is still parametrically much smaller than the special (single) mode contribution.
\footnote{Note that
we are assuming $M_N^{ \rm UV } \ll M_{ \rm Pl }$ here, although the hierarchy here need only be an order of magnitude or so for the
5D mass basis results (for the special mode) to be valid.} Another crucial contribution must come from somewhere else.
%of  canceling against the special (single) mode contribution, i.e., 
%our ``optimism'' above is dashed (or hope above is not quite realized)!

\vspace{0.1in}

\noindent {\bf Modes {\em near} special mode} 

\vspace{0.1in}

What remains to be considered for the resolution of the above ``discrepancy'' is to take into account a ``threshold'' effect at the scale of the special mode, i.e., include the contribution to the SM neutrino mass from the paired special modes.
Indeed, we find that the
modes just above and below the special mode are also ``special'' (even if {\em paired}) in the sense that 
the naive 
extrapolation for their properties 
%(to be specific/in particular, mass and coupling splitting) 
from
the formulae for low-lying modes is simply invalid. 
For example, 1st line of Eq.~(\ref{5Dmass_split}) would give mass splitting $\sim \left( M_N^{ \rm UV } / M_{ \rm Pl } \right)^{ -2 \; c_N - 1 } \times$ TeV, i.e., $\ll$ TeV, by setting $m_n \sim M_N^{ \rm UV }$, but actually we find that 
it is $\sim$ TeV (see appendix \ref{sec:app_masses} and \ref{sec:app_couplings}).
And, the Dirac mass with the SM doublet neutrino for these modes (at the leading order) is similar to that of the special, single mode, i.e.,
Eq.~(\ref{mDspecial2}) (again, 
as dictated by all these profiles being peaked near the TeV brane). 
Thus, for each such pair, the contribution to the SM neutrino mass of the mass splitting {\em by itself} (i.e., setting couplings to be exactly {\em degenerate}:
we will return to the splitting in couplings momentarily!)
is
\bea
m_{ \nu }^{ \rm special, one-pair } \; (\hbox{mass splitting {\em only}}) & \sim &  v^2 \left( \frac{ \hbox{TeV} }{ M_{ \rm Pl } } \right)^{ 2 \left( c_L - \frac{1}{2}  \right) } 
\frac{ \Delta M_{ \rm special } }{ m_{ \rm special }^2 } \nonumber \\
& \sim & v^2 \left( \frac{ \hbox{TeV} }{ M_{ \rm Pl } } \right)^{ 2 \left( c_L - \frac{1}{2}  \right) }  
\frac{ \rm TeV }{ {M_N^{ \rm UV }}^2 }
\label{mnu_special_pair_masssplit}
\eea
Now, the 
number of such special, {\em paired} modes is approximately given by (see appendix \ref{sec:app_masses})
\bea
\eta_{\rm special, paired } & \sim & j_{<-1/2} \left( \frac{ M_{ \rm Pl } }{ \rm TeV } \right)
\left( \frac{ M_N^{ \rm UV } }{ M_{ \rm Pl } } \right)^{ -2 \; c_N } 
\eea
with 
\bea
j_{<-1/2} \sim \frac{2\pi (-1/2-c_N)^{1-2c_N} \tan(c_N \pi)}{\Gamma^2 (-c_N+1/2)}
\eea
Upon 
summing Eq.~(\ref{mnu_special_pair_masssplit}) over these special modes, we then get
\bea
m_{ \nu }^{ \rm special, all-pairs } \; (\hbox{mass splitting {\em only}})  & \sim & \frac{v^2}{M_N^{\rm UV}} \left( \frac{ \hbox{TeV} }{ M_{ \rm Pl } } \right)^{ 2 \left( c_L - \frac{1}{2}  \right) } 
\left( \frac{ M_N^{ \rm UV } }{ M_{ \rm Pl } } \right)^{- 2 \; c_N - 1 }  
\eea
i.e., same size as the sum over {\em non}-special modes (cut-off as above), see Eqs.~(\ref{mnu_non_special}) and (\ref{mnu_special_single}),
so that this is 
still not enough to cancel the excessive contribution of the special, single mode.

However, what ``saves the day'' is that the effect of the {\em coupling} splitting for these paired-special modes is actually larger, i.e., dominates over the mass splitting.
%(cf.~for {\em non}-special modes both for $c_N < -1/2$ and $c_N > -1/2$ or for modes near special one for $c_N > -1/2$).
%
In detail, the {\em relative} splitting in coupling (and hence in Dirac mass term with the SM doublet neutrino) is given by (see appendix \ref{sec:app_couplings})
\bea
\delta_{ \rm coupling }^{ \rm special } & \sim & \left( \frac{ \rm TeV }{ M_{ \rm Pl } } \right)
\left( \frac{ M_N^{ \rm UV } }{ M_{ \rm Pl } } \right)^{ 2 \; c_N }
\eea
so that contribution to the SM neutrino mass from this effect for {\em each} pair is: 
\begin{adjustwidth}{-0.5cm}{}
\bea
m_{ \nu }^{ \rm special, one-pair } \; (\hbox{coupling splitting}) & \sim & v^2 
\left( \frac{ \hbox{TeV} }{ M_{ \rm Pl } } \right)^{ 2 \left( c_L - \frac{1}{2}  \right) } 
\frac{ \delta_{\rm coupling }^{ \rm special } }{ M_N^{ \rm UV } } \nonumber \\
& \sim & v^2 \left( \frac{ \hbox{TeV} }{ M_{ \rm Pl } } \right)^{ 2 \left( c_L - \frac{1}{2}  \right) } \frac{ \rm TeV }{ M_{ \rm Pl }^2 } 
\left( \frac{ M_N^{ \rm UV } }{ M_{ \rm Pl } } \right)^{ 2 \; c_N - 1 } 
\eea
\end{adjustwidth}
clearly larger than the mass splitting effect of Eq.~(\ref{mnu_special_pair_masssplit}).
And, summing over special mode pairs, gives (we multiply the previous result by $\eta_{\rm special, paired }$):
\bea
m_{ \nu }^{\rm special, all-pairs } \; (\hbox{coupling splitting}) & \sim & 
\frac{ v^2 }{ M_N^{ \rm UV } }
\left( \frac{ \hbox{TeV} }{ M_{ \rm Pl } } \right)^{ 2 \left( c_L - \frac{1}{2}  \right) }.
\eea
which is indeed larger than sum of non-special modes (cut-off at special mode mass) in Eq.~(\ref{mnu_non_special}).
 Importantly, 
the above collective effect is parametrically {\em comparable} to that of the special mode by itself in Eq.~(\ref{mnu_special_single}). So the two ``special'' contributions -- single and paired (again, with mass $\sim M_N^{ \rm UV }$) -- {\em can} cancel each other to a large extent! 

We thus conjecture that this is precisely what happens: it is the sum over {\em all} modes -- special (paired and single) and ordinary below it
-- which can reproduce the KK basis result for $c_N < -1/2$.

\vspace{0.1in}

\noindent {\bf Modes (much) {\em above} special mode} 

\vspace{0.1in}

For the sake of completeness, especially given the
``divergence'' in the bottom-up approach, we should carefully estimate the effect from modes (much) above special one: we indeed find this to be convergent and negligible.
% (as compared to the effects studied above).
In more detail, an analysis similar to that performed for modes below special one shows that the mass {\em splitting} in each pair for $M_{ \rm Pl } \gg 
m_n \gg M_N^{ \rm UV }$ is given by
\bea
\Delta m \; \hbox{for} \; m_n \gg M_N^{ \rm UV } & \sim & \hbox{TeV} \left( \frac{ M_N^{ \rm UV } }{ M_{ \rm Pl } } \right)
\left( \frac{ m_n }{ M_{ \rm Pl } } \right)^{ - 2 \; c_N - 2 } 
\eea
whereas the Dirac mass term with the SM doublet neutrino is similar to the other mass eigenstates, i.e., Eq.~(\ref{mDspecial2}).
So, the contribution of each such {\em pair} to the SM neutrino mass is given by
\bea
m^{ \rm pair } _{ \nu } & \sim & \left( m_D^{ \rm special, single } \right)^2 \left( \frac{ \hbox{TeV} }{ m_n^2 } \right) 
\left( \frac{ M_N^{ \rm UV } }{ M_{ \rm Pl } } \right)
\left( \frac{ m_n }{ M_{ \rm Pl } } \right)^{ - 2 \; c_N - 2 } 
\eea
Thus, we see that the sum over these modes (setting $m_n \sim n \times$TeV as usual) is convergent (as long as $c_N > -3/2$). Their total contribution is much smaller than the (summed) 
contribution of the {\em low}-lying modes [see Eq.~(\ref{mnu_non_special})] by $\sim \hbox{TeV} / M_N^{ \rm UV }$.

\vspace{0.1in}

%
%To be really 
%
%careful
%
%safe, we should 
%consider carefully modes (much) above special one: indeed, we find their contribution to be negligible}.
%

%
%We are then led to argue that this sum of modes (up to roughly the special mode) ``must'' mostly cancels %against the special mode contribution
%-- {\em both} being overly large -- in 
%such a manner that we are 
%leftover with the (much smaller relative to each) effect of just the lowest/TeV-scale modes; in this manner, 
%it is the sum over {\em all} modes -- special and ordinary -- which 
%
%finally
%
%reproduces the KK basis result even for $c_N > 1/2$.
%

\section{CFT interpretation}
\label{CFT}

%%%%%%%%%%%%%%%%%
% moved from earlier introduction
%%%%%%%%%%%%%%%%%

%As outlined above, in this paper, we also provide the {\bf CFT picture} of the above 5D model.

Let us start by reminding the reader the CFT interpretation of bulk {\em charged} SM fermions. In this case a massless chiral external fermion (often called ``elementary'') is coupled (at the UV cut-off) to a CFT fermionic operator: the scaling dimension of this operator (and hence the size of this coupling in the IR, up on RGE from UV cut-off) is related to the 5D mass parameter. The mass eigenstates, which correspond to the zero and KK modes of the 5D model, are actually {\em admixtures} of the external fermion and composite fermions interpolated by the CFT operator.

For the case of the singlet neutrino at hand, there is an additional feature: the external fermion (denoted by $N_R$) has a Majorana mass term whose size can be close to the UV cut-off. Denoting by ${\cal O}_N$
the CFT operator to which $N_R$ couples, the UV Lagrangian contains
\bea
\mathcal{L} = \mathcal{L_{\rm CFT}} + \lambda \overline{N_R} \mathcal{O}_N + \frac{1}{2} M_N^{\rm bare} N_R^2
\eea
where we are using the convention that the {\em engineering} dimension of ${\cal O}_N$ is 5/2 so that the coupling $\lambda$ is dimension{\em less}. We take the natural size of bare Majorana mass $M_N^{\rm bare} \lesssim M_{\rm Pl}$.
The composite operator $\mathcal{O}_N$ actually interpolates {\em left}-handed composite fermionic states.
These composites form {\em Dirac} states, with masses being quantized in units of $\sim$ TeV and 
with their RH partners originating from a {\em different} operator (which will not concern us here). 
Due to the above coupling, there is mixing between $N_R$ and CFT composites so that the
basis defined by the external $N_R$ and the CFT composites is 
not quite the mass basis of the 5D model that we discussed above, not the KK basis of 5D model. We dub it ``CFT'' basis.
This provides yet another angle on the seesaw mechanism, allowing us to obtain quick estimates as we discuss below.

\vspace{0.1in}

\noindent (i) {\Large $\big[ {\cal O}_N \big] < 5/2$ or $c_N > -1/2$}

\vspace{0.1in}

The coupling $\overline{ N_R} {\cal O}_N$ is {\em relevant} when the scaling dimension of operator, denoted by $\big[ {\cal O}_N \big]$, is less than 5/2.
In this scenario, the (CFT $+ N_R$) theory
flows to a new fixed point and we assume it is reached rather rapidly, just below the UV cut-off $\sim M_{ \rm Pl }$. At the fixed point, $N_R$ effectively has a scaling dimension of $\left( 
4 - \big[ {\cal O}_N \big] \right)$ so that the net coupling $\overline{ N_R} {\cal O}_N$
has a scaling dimension of 4, as appropriate for a fixed point behaviour \cite{Contino:2004vy}. 

\vspace{0.1in}

\noindent 
{\bf Mass of  $N_R$}

\vspace{0.1in}

The mass {\em term} for $N_R$ can be significantly renormalized (actually reduced) compared to its bare value.
The RG running is dominantly dictated by anomalous dimension of the operator $N_R^2$
%, $\gamma_{N_R^2} = 2 \big[ N_R \big] - 2 \cdot \frac{3}{2}$, 
and we find
\bea
M_N \left( \mu \right) & \sim & M_N^{ \rm bare } \left( \frac{ \mu }{ M_{ \rm Pl } } \right)^{ 5 - 2 \big[ {\cal O}_N \big] }, 
\; \hbox{for} \; \big[ {\cal O}_N \big] < 5/2
%
%\label{}
%
\eea
where we assumed the large-$N$ limit \footnote{Here, ``$N$'' denotes (roughly) the number of fundamental degrees of freedom in the CFT, which is
not to be confused with the singlet fermion {\em field} $N$!} 
%(as is relevant for the AdS/CFT duality to apply to this 5D model) 
in taking scaling dimension of $N_R^2$ field to be twice that of $N_R$ (and we have set the engineering dimension of $N_R$ to be $3/2$).

It is natural to assume that the ``physical mass'' for $N_R$ (denoted by $M^{ \rm phy }_N$) is given by the value of $\mu$ where the renormalized mass term 
becomes comparable to $\mu$ itself ,
\bea
M^{ \rm phy }_N  & \sim & M_N^{ \rm bare } \left( \frac{ M^{ \rm phy }_N }{ M_{ \rm Pl } } \right)^{ 5 - 2 \big[ {\cal O}_N \big] }
\label{MNphycond}.
\eea
Solving for $M^{ \rm phy }_N$ gives
\bea
M^{ \rm phy }_N  & \sim & M_N^{ \rm bare } \left( \frac{ M_N^{ \rm bare } }{ M_{ \rm Pl } } \right)^{ \frac{1}{ 2 \big[ {\cal O}_N \big] - 4 } - 1 }.
\label{MNspecial1CFT}
\eea
Note that the exponent on RHS in equation just above is indeed $>0$ for $\big[ {\cal O}_N \big] < 5/2$ so that $M_N^{ \rm phy } < M_N^{ \rm bare }$. 
Of course, $N_R$ mixes with CFT states (that is why we used quotes while calling $M^{ \rm phy }_N$ a mass), but it is clear that there will be a resultant 
mass eigenstate with significant admixture of $N_R$, which thus has a mass roughly given by the renormalized $N_R$ mass term.

When matching to the 5D results, we use the standard AdS/CFT ``dictionary'': first, we can relate
$\big[ {\cal O}_N \big]$ to the 5D mass of $N$, namely, $\big[ {\cal O}_N \big] = 2 - c_N$.
Thus, it is $c_N > -1/2$ which corresponds to the relevant $\overline{ N_R} {\cal O}_N$ coupling assumed above.
And, $M_N^{ \rm bare }$ in the CFT picture is dual to the Majorana mass term on the UV brane, $M_N^{ UV }$.
Plugging in the {\em parameters} into Eq.~(\ref{MNspecial1CFT}), we recover the mass of the special mode in Eq.~(\ref{MNspecial1}).

\vspace{0.1in}

\noindent 
{\bf {\em Low}-lying modes}

\vspace{0.1in}

Effectively integrating out $N_R$ at the scale $M^{ \rm phy }_N$ gives rise to the composite operator ${\cal O}_N^2$, thus feeding lepton-number violation into the CFT sector:
\bea
\Delta {\cal L}_{ CFT } & \sim &  \lambda \overline{N_R} \mathcal{O}_N + \frac{1}{2} M_N^{\rm phy} N_R^2  \nonumber  \\
& \rightarrow & \frac{\lambda^2}{ M^{ \rm phy }_N } {\cal O}_N^2, \; \hbox{renormalized at} \; M^{ \rm phy }_N \label{DeltaCFTbare}
\eea
where $\Delta {\cal L}_{ CFT }$ denotes perturbation to the CFT Lagrangian.
RG evolving this to the $\sim$ TeV scale (as before, we use $\big[ {\cal O}_N^2 \big] = 2 \times \big[ {\cal O}_N \big]$, similarly for the engineering dimensions), 
%
%Higgs is created/born
%EW is broken
%
where composite Higgs is interpolated by the {\em product} of ${\cal O}_N$ and $O_L$ (latter being the doublet operator)\footnote{Note that had we taken Higgs field also to be in the bulk (but with profile of its VEV/SM Higgs boson to be peaked near TeV brane), then we would have a single trace, finite/low scaling dimension CFT operator, ${\cal O}_H$ which can also interpolate the composite Higgs. Instead, we assumed here -- mostly for simplicity -- that Higgs is strictly localized on the TeV brane which 
implies that there is no such ``Higgs'' operator at higher than $\sim$ TeV energies.},
we get
\bea
\Delta {\cal L}_{CFT} & \sim & \frac{\lambda^2}{ M^{ \rm phy }_N } \left( \frac{ \rm TeV }{ M^{ \rm phy }_N }  \right)^{ 2 \big[ {\cal O}_N \big] - 5 } {\cal O}_N^2 , \; \hbox{renormalized at TeV} \nonumber \\
& \sim &  \frac{\lambda^2}{ M^{ \rm bare }_N } \left( \frac{ \rm TeV }{ M_{ \rm Pl } }  \right)^{ 2 \big[ {\cal O}_N \big] - 5 } {\cal O}_N ^2 \label{ON^2TeV}\\
& \sim & \frac{\lambda^2}{\rm TeV } \left( \frac{ \rm TeV }{ M^{ \rm phy }_N }  \right)^{ 2 \left( \big[ {\cal O}_N \big] - 2 \right)} {\cal O}_N ^2 \nonumber
\eea
using Eq.~(\ref{MNspecial1CFT}) in 2nd line above.

Based on the above RG scaling and the requirement of stability of the system, we find that there is a {\em lower} limit on $\big[ {\cal O}_N \big]$.
Suppose the dimensionless coefficient of the Lagrangian term in 
Eq.~(\ref{DeltaCFTbare}) is $\sim O(1)$, i.e., it starts being a ``borderline'' perturbation to the CFT.
However, even with this assumption about the initial condition,
as can be seen from the last line of Eq(\ref{ON^2TeV}), in the IR, it will always be a {\em genuine} perturbation, i.e., the coefficient (in units of the corresponding RGE scale) $\ll 1$, {\em as long as $\big[ {\cal O}_N \big] > 2$ so that ${\cal O}_N^2$ is an {\em ir}relevant operator}.
% (i.e., has scaling dimension $>4$)}.
In 5D we thus require $c_N<0$, which is what we assumed in our calculations.\footnote{In other words, for the case
$\big[ {\cal O}_N \big] < 2$, we see that ${\cal O}^2_N$ is a relevant operator. 
The ``problem'' with this scenario 
%
%$\big[ {\cal O}_N \big] < 2$ 
%
is that, even if the coefficient in Eq.~(\ref{DeltaCFTbare}) is {\em smaller} than 1, it
will become (again, in appropriate units) {\em larger } than 
$\sim O(1)$ at an RG  scale which is (possibly much) {\em above} $\sim$ TeV, i.e., 
there is a danger that scale invariance is then broken at that scale.} 
%(which would of course be undesirable!).}
%

\vspace{0.1in}

\noindent 
{\bf SM neutrino mass}

\vspace{0.1in}

Interpreting Eq.~(\ref{ON^2TeV}) as the main source for lepton-number violation, introducing a factor of
$\sim \left( \hbox{TeV} / M_{ \rm Pl } \right)^{ 2 \big[ {\cal O}_L \big] - 5 }$ for the (square of) coupling of doublet lepton neutrino to the CFT in the IR \cite{Contino:2004vy} \footnote{Recall that, as discussed in section \ref{KK}, $c_L \sim 0.6$ reproduces charged lepton masses and this corresponds to $\big[ \mathcal{O}_L \big] > 5/2$, i.e. irrelevant coupling.} and Higgs VEV for EWSB, we estimate the SM neutrino mass:
\bea
m_{ \nu } & \sim & \frac{ v^2 }{ M^{ \rm bare }_N } \left( \frac{ \rm TeV }{ M_{ \rm Pl } }  \right)^{ 2 \left( \big[ {\cal O}_N \big] + \big[ {\cal O}_L \big] - 5 \right) } 
\label{mnuCFT}
\eea
Upon translating to the 5D parameters, we again get agreement for another physical observable, namely, the SM neutrino mass in Eq.~(\ref{mnuCFT}) is similar to the result obtained using the 5D calculation in Eq.~(\ref{mnu5D}).
%
%even if not as short as in KK basis!

In the CFT picture, we can also think 
%(like we did in the 5D model) 
in terms of the SM neutrino mass actually arising from {\em exchange} of heavy SM singlet particles.
% (``anatomy'' of SM neutrino mass generation if you will!). 
%
The point is that the above lepton-number violating perturbation ${\cal O}_N^2$ to the CFT 
%(namely ${\cal O}_N^2$ from integrating out the external $N_R$, with a Majorana mass) 
will induce small Majorana mass terms and lepton-number violating couplings to the Higgs for the entire tower of CFT {\em composites}, 
%(as anticipated in the summary above), 
which of course are SM singlets and Dirac.
% (to begin with).
%
In more detail, using Eq.~(\ref{ON^2TeV}), it is rather straightforward to estimate this effect for the {\em lightest} TeV-scale composites. For example, the mass splitting is of order:
\bea
%
%\frac{ 
%
\Delta M \; \hbox{from} \; {\cal O}_N^2 
%
%}{ \rm TeV } 
%
& \sim &  \frac{ \hbox{ TeV}^2 }{ M^{ \rm bare }_N } \left( \frac{ \rm TeV }{ M_{ \rm Pl } }  \right)^{ 2 \big[ {\cal O}_N \big] - 5 }
\label{DeltaMCFT}
\eea
After diagonalizing these mass terms 
%(again, which mix CFT (again, which are {\em a priori} Dirac) composites, including of different levels),
it is clear that we will obtain pairs of (almost) degenerate {\em Majorana} modes with mass splitting as in Eq.(\ref{DeltaMCFT}), and this is what we found in the 5D mass basis calculation.
Speaking more quantitatively, relating the scaling dimension of ${\cal O}_N$ to $c_N$ and identifying $M^{ \rm bare }_N$ with $M_N^{ UV }$, we see that this Majorana mass term 
%(and hence splitting within quasi/mostly-Dirac pair of CFT composite) 
has the same size as in Eq.~(\ref{5Dmass_split}) of the 5D calculation.

\begin{figure}
\begin{adjustwidth}{-0cm}{-0cm}
\centering
\mbox{\includegraphics[height=50mm]{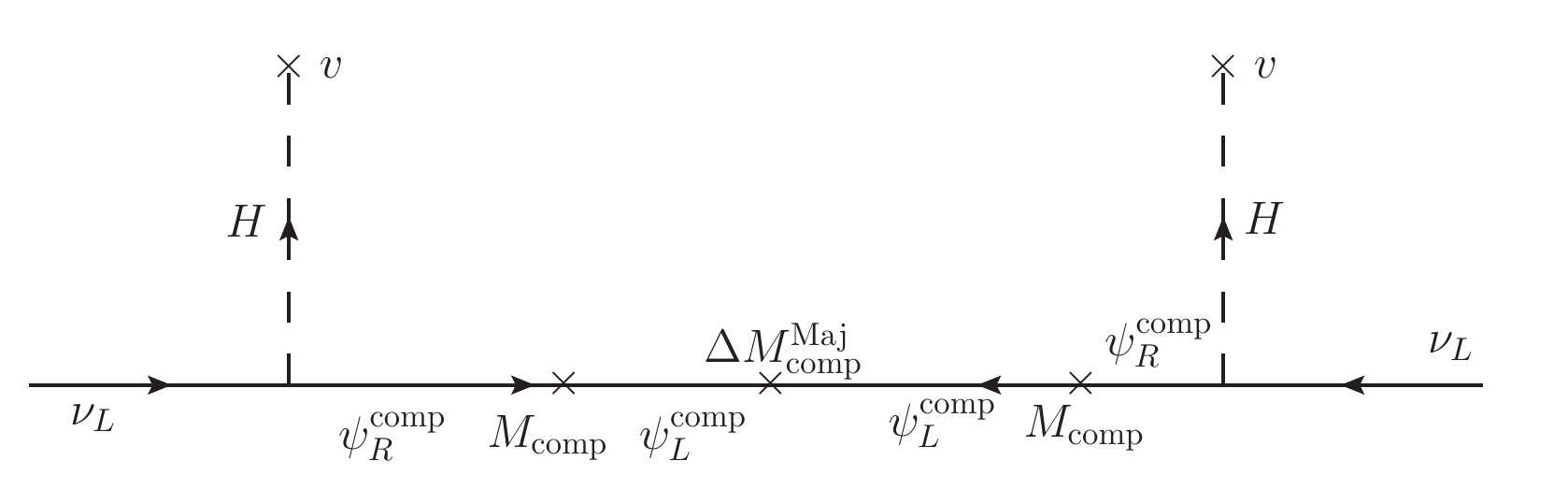}}
\caption{The SM neutrino mass generated by exchange of one composite state in 
the CFT basis, labelled $\psi^{ \rm comp }$
with Dirac mass $M_{ \rm comp }$ and Majorana mass term 
$\Delta M_{ \rm comp }^{ \rm Maj}$. The chirality structure is to be %
contrasted to that in Fig.~\ref{nu_mass_KK} for the KK basis.}
\label{nu_mass_CFT}
\end{adjustwidth}
\end{figure}

%
%Then/Using/
%
Armed with these Majorana mass terms for the TeV-scale composites, it is rather straightforward to show
%(along the lines of what was done in the 5D mass basis) 
that the contribution to the SM neutrino mass from the exchange of the low-lying resonances provides an order one contribution to the SM neutrino mass. Interestingly,
\begin{itemize}
\item
the Majorana mass term is for the {\em left}-handed composites (again, interpolated by ${\cal O}_N$),
whereas coupling to the Higgs is for the R chirality so that
we do not encounter any propagator suppression in the exchange of TeV-scale composites (as opposed to the KK basis), see Fig.~\ref{nu_mass_CFT}.

\end{itemize}

We see from Eq.~(\ref{DeltaMCFT}) that 
$\Delta M \ll$ TeV,
as long as $\big[ {\cal O}_N \big] > 2$ (as we assumed above for stability).
Also, just to make the point from the earlier summary more explicitly, 
%
%\ka{once again, I am Ok with removing the ``bullet'' from the following!}, we see that
for $N_R \overline{ {\cal O}_N}$ coupling being close to marginal 
%
%operator 
%
(i.e., $\big[ {\cal O}_N \big] \sim 5/2$)\footnote{deviating from marginality 
does not really change the point which follows}, we get $\Delta M \sim \hbox{TeV}^2 / M_N^{ \rm bare }$, i.e., Majorana mass term for CFT composites is {\em naturally} suppressed because it 
sort of manifests 
%(or is actually/physically the result of) 
a 
%(type I/high-scale) 
``seesaw'', 
with $\sim$ TeV in numerator being (roughly) Dirac mass term between $N_R$ and (TeV-scale) CFT composite and $M_N^{ \rm bare }$ being Majorana mass for $N_R$ which is heavy and integrated out: 
of course, the ``difference'' from usual seesaw for SM neutrino mass is that here CFT composite also has a {\em Dirac} mass 
$\sim$ TeV (with another composite).

In addition, it is worth mentioning that the Majorana mass term 
%(for $\sim$ TeV mass Dirac states, in this case CFT composites) 
which is needed for obtaining SM neutrino mass [i.e., $\sim O(0.1)$ eV] from exchange of these TeV-mass modes is actually $\sim $ keV, i.e., several orders of magnitude {\em larger} than simply $\sim \hbox{TeV}^2 / M_{ \rm Pl } \sim$ meV that we would have gotten for 
the $N_R-{\cal O}_N$ coupling being marginal 
(as indicated above) and $M_N^{ \rm bare } \sim M_{ \rm Pl }$. Yet, here we have an interesting option:
%/it is enough to choose 
%
%all that we need/suffices
%
\begin{itemize}

\item
for $[{\cal O}_N]\lesssim5/2$, i.e., a {\em slightly} relevant coupling of $N_R$ to CFT operator, naturally gives the requisite size of Majorana mass term for TeV-mass Dirac composites [as seen from Eq.~(\ref{DeltaMCFT})]: 
the crucial point being that 
a small deviation from marginality for the above coupling is ``enhanced'' by RGE over the large energy range.
%
%to give the desired Majorana mass
%(term) for the TeV-scale composites.
%
%(large) suppression for $N_R$.
%

\end{itemize}

%
%
%\begin{figure}

%\vspace{-10cm}

%\hspace{-2.5cm}

%\includegraphics{nu_mass_CFT}

%\vspace{-15cm}

%\caption{The SM neutrino mass generated by exchange of one composite state in 
%the CFT basis, labelled $\psi^{ \rm comp }$
%with Dirac mass $M_{ \rm comp }$ and Majorana mass term 
%$\Delta M_{ \rm comp }^{ \rm Maj. }$. The chirality structure is to be %
%contrasted to that in Fig.~\ref{nu_mass_KK} for the KK basis.}
%\label{nu_mass_CFT}
%\end{figure}
%

Finally, we have seen that the TeV scale composites provide an important contribution to the SM neutrino mass. On the other hand, while $N_R$ is {\em crucial} in introducing the seed of lepton-number violation in the CFT via ${\cal O}_N^2$, 
%(again, it is {\em not} precisely the mass eigenstates being exchanged here, although it is close to that picture!), 
$N_R$ itself does {\em not} directly couple to the Higgs. So, we learn that
\begin{itemize}

\item
there is no {\em additional} contribution to the SM neutrino mass from $N_R$ exchange per se, even though $N_R$ has a Majorana mass: what is missing is the coupling to the Higgs.

\end{itemize}

%
%And, from CFT picture/basis, it is clear that estimate for Majorana mass term induced for singlet CFT composites 
%hence
%the final SM neutrino mass has the same dependence on $c_N$/functional form associated with scaling dimension of N
%(in other words, $c_N$ in 5D picture), in agreement with KK basis result of Eq.~(\ref{}) and Eq.~(\ref{}) obtained using 
%mass basis in 5D
%

\vspace{0.1in}

\noindent (ii) {\Large $\big[ {\cal O}_N \big] > 5/2$ or $c_N < -1/2$}

\vspace{0.1in}

The CFT picture for $c_N < -1/2$ should then be easy to go through; to begin with, 
the usual translation dictionary  implies $\big[ {\cal O}_N \big] > 5/2$ so that
coupling $\overline{ N_R } {\cal O}_N$ is now {\em ir}relevant. 
%(cf.~the relevant coupling for $c > -1/2$ discussed earlier). 
%
Thus, it is clear that the 
%
%physical 
%
mass {\em term} for $N_R$ is roughly the size of the Majorana mass term at the UV cut-off itself, i.e., there is negligible renormalization for it.  
Moreover, 
as before, we can argue that in spite of the mixing of $N_R$ with CFT composites there will be an ``$N_R$ state'' whose physical mass is not significantly modified relative to the $N_R$ mass term above, i.e., 
\bea
M_N^{ \rm phy } & \sim & M_N^{ \rm bare }, \; \hbox{for} \; \big[ {\cal O}_N \big] > 5/2
\eea
which is of course in agreement with the 5D single-special mode mass [see Eq.~(\ref{MNspecial2})] for this case.

We can integrate out $N_R$ as before, except that this is now done at $M_N^{ \rm bare }$, %cf.~somewhat  smaller scale for $c_N < 1/2$: see Eq.~(\ref{MNspecial1CFT}).
RG flowing from this scale to $\sim$ TeV, it is easy to see that the $c_N < -1/2$ (or $\big[ {\cal O}_N \big] > 5/2$) case actually gives similar form for the coefficient of ${\cal O}_N^2$ operator as $c_N > -1/2$ (or $\big[ {\cal O}_N < 5/2 \big]$) that we discussed earlier; this happens mainly because the only assumption we made earlier for this purpose 
about $\big[ {\cal O}_N \big]$ was that it is larger than 2, which is certainly the case for $c_N < -1/2$.
Hence, the SM neutrino mass for $c_N > -1/2$ in the CFT picture is also given by Eq.~(\ref{mnuCFT}) and, in turn, agrees with the 5D result in Eq.~(\ref{mnu5D}).
Again, the SM neutrino mass originates only from CFT composites exchange [with Majorana mass terms for $\sim$ {\em TeV}-scale composites given as before:
see Eq.~{(\ref{DeltaMCFT})], since external 
$N_R$ does not couple to the Higgs in this basis.

\vspace{0.1in}

\noindent {\bf Contribution to the SM neutrino mass from special modes for $c_N < -1/2$} 

\vspace{0.1in}

%
%What about the 
%
Using the CFT picture, can we understand the {\em un}expectedly large contribution to the SM neutrino mass  of the {\em special} mode in {\em mass} basis found in the 5D calculation for
$c_N < -1/2$? 
Note that this CFT basis 
%[where we chose to work in terms of $N_R$ and CFT composites (separately) as the states] 
is {\em not} exactly the mass basis. Thus, first of all, there is no obvious ``contradiction'' between $N_R$ exchange in CFT picture {\em not} (directly) contributing to the SM neutrino mass with the fact that, in the mass basis, 
%
%(obtained using 5D calculation), 
%
the special mode gives a large contribution 
%(which turns out to be {\em too} large!) 
to the SM neutrino mass, in turn, from its {\em un}suppressed\footnote{as usual,
apart from being possibly small due to choice of $c_L$ or $\big[ {\cal O}_L \big]$.} 
coupling to the Higgs. The point is that 
\begin{itemize}

\item
the special mode of the 5D model 
%(in mass basis) 
would in the CFT picture correspond to an {\em admixture} of $N_R$ and CFT composites and the latter component of it does couple 
%-- again, unsuppressed -- 
to another composite, i.e., the Higgs: 
first of all, this implies
% (based on this {\em CFT} picture) 
that the special mode {\em will} couple to the Higgs (as we found in the 5D calculation).

\end{itemize}
Thus, the ``origin'' of the special mode and how it contributes to the SM neutrino mass 
%(that too, a type I seesaw) 
is clear from the CFT perspective.

%\footnote{\ka{I modified the following: it's too long. So, I am open to shortening it!} It is interesting that an attempt to provide an explanation/UV-completion for the small Majorana mass term for Dirac states of the {\em usual/4D} (purely) inverse seesaw model can actually "lead" us (as follows) to the CFT picture discussed above (and thus its dual 5D model). One could start with the 4D/basic inverse seesaw model which contains a Dirac field made of two chiralities $N$ and $S$, with $N$ being the one coupling to doublet neutrino/Higgs (aka "RH neutrino" so that $S$ is to be thought of as LH). We can essentially remove the {\em bare} (small) Majorana mass term for $S$. Instead, we add a super-heavy Majorana field (call it $\tilde{N}$) and then give it a mass mixing term (of size $\sim$ TeV or smaller) with $S$. Integrating out this $\tilde{N}$ would then (naturally) generate {\em small} Majorana mass term for the {\em left} chirality $S$. This is of course/basically the CFT picture, with  $N$, $S$ (of the usual 4D inverse seesaw model) forming TeV-mass (Dirac) CFT composite and the added $\tilde{N}$ being the $N_R$ external to the CFT. So, it is somewhat amusing that such a (improvement) procedure/process applied to the usual inverse seesaw model leads to the {\em addition} (to the inverse seesaw piece) of a {\em type I} seesaw contribution (in the mass basis for all these modes), i.e., from the (very) {\em heavy} $\tilde{N}$ (of course mixed with the original Dirac state), which corresponds to just the ``special'' mode in the 5D model!}
%
%language
%
But, the main question still remains, namely, how come special mode's coupling to the Higgs %(obtained as above) 
is so {\em large}, given that the coupling between $N_R$ and the CFT is {\em small} for the case $\big[ {\cal O}_N \big] > 5/2$ ?
%(i.e., {\em ir}relevant coupling between $N_R$ and CFT operator)?!
%
The answer to this puzzle is the following. There is a whole tower of CFT composites (from $\sim$ TeV to $M_{ \rm Pl }$)
with which $N_R$ mixes. In particular there are many composites which have mass $\sim M_N^{ \rm phy}$. Therefore, even the small off-diagonal mass terms between $N_R$ and these CFT composite states can result in {\em large} mixing {\em angles}. This mixing -- even if it is close to maximal -- 
does not really change the {\em physical} mass of $N_R$ from the mass {\em term} for $N_R$. Conversely, the coupling can be modified significantly. In particular, we see that the special mode can acquire large coupling to the Higgs by ``piggy-back riding'' on the coupling of its sizable admixture of (almost) degenerate CFT composites. Schematically, we have:
\bea
\hbox{special mode constitution}
& \propto & N_R + a \psi^{ \rm near} + \epsilon \psi^{ \rm far } 
\eea
where $\psi^{ \rm near }$ denotes (collectively) the CFT composites with mass close to $M_N^{ \rm bare }$ ($\psi^{ \rm far }$ denoting rest of the CFT tower) and $a$ is $\sim O(1)$ mixing angle, whereas $\epsilon \ll 1$. %, i.e., following the naive ``rules of the game''). 
Thus, in the end, the special mode has $O(1)$ coupling to the Higgs.
%
%so does final result for SM neutrino mass: 
%
%
%as already mentioned, this contribution same form for $c_N > 1/2$ as $c_N < 1/2$ even in (actually naturally in) CFT picture...
%

Note that 
%(in principle) 
a similar argument applies to the case $c_N > -1/2$ or $\big[ {\cal O}_N \big] < 5/2$ studied earlier. 
However, there the mixing mass term, i.e.,  $\delta m_{ N_R-\rm CFT}$, can be sizable to begin with, given that the coupling between $N_R$ and CFT operator is {\em relevant}.
Thus, the closeness in mass of some CFT composites with $N_R$ has less of an {\em additional} impact as compared to the case $c_N < -1/2$ discussed above, i.e., there is really no ``(further) enhancement'' of the mixing effect here.
Also, the special mode 
-- being too heavy compared to would-be zero mode -- does not contribute significantly to the SM neutrino mass, {\em even if} its coupling
to the Higgs is taken to be unsuppressed\footnote{cf. for $c_N < -1/2$, where the (unexpectedly) large coupling to Higgs changed the game drastically!} (and similarly for modes around it). 
Overall, 
that is why this issue of taking into account mixing between $N_R$ and CFT composites 
is not really significant for $c_N > -1/2$, i.e., we do not expect to find (and indeed did not in the 5D calculation) any ``surprises'' here.

\vspace{0.1in}

\noindent {\bf ``Universal'' dependence on $c_N$ of the SM neutrino mass} 

\vspace{0.1in}

Moreover,
%
%Thus, 
%
The CFT picture leads to a simple ``understanding'' of why the dependence on $c_N$ in the formula for SM neutrino mass obtained from 5D calculation is the same for $c_N < -1/2$ and $c_N > -1/2$ [see Eq~(\ref{mnu5D})]; as have been discussed in section \ref{KK}, this looked somewhat of a coincidence in the KK basis.
The SM neutrino mass in the CFT picture is essentially dictated by the lepton-number violating effect in the CFT sector, i.e., the  coefficient of the operator ${\cal O}_N^2$ renormalized at $\sim$ TeV scale\footnote{in the anatomical language, this operator first leads to Majorana mass terms for the CFT singlet composites, whose exchange then generates the SM neutrino mass.}.
In turn, this is determined by $\big[ {\cal O}_N \big]$, the scaling dimension of ${\cal O}_N$ (that of ${\cal O}_N^2$ being twice in the
large-$N$ limit). 
%
%\begin{itemize}
%
%\item
The key observation is that, as long as $\big[ {\cal O}_N^2 \big] > 4$ (thus $\big[ {\cal O}_N \big] > 2$)
%i.e., ${\cal O}_N^2$ is an irrelevant operator,
the RG flow of coefficient of 
${\cal O}_N^2$ (down to the TeV scale) has a similar dependence on $\big[ {\cal O}_N \big]$. This range of $\big[ {\cal O}_N \big]$ corresponds to $c_N < 0$, whether $c_N < -1/2$ or $c_N > -1/2$.
%
%\end{itemize}
%
Hence, we do not expect any qualitative change in the formula for the SM neutrino mass as we cross the $c_N = -1/2$ ``threshold'': 
again, while this marks the transition of the coupling $\overline{ N_R } {\cal O}_N$ from relevant to irrelevant, it is $\big[ {\cal O}^2_N \big]$ which matters for the bottomline SM neutrino mass and this operator stays irrelevant throughout this range of $c_N$.

\section{Conclusions and outlook}
\label{conclude}

We studied a simple warped 5D scenario that accommodates the SM neutrino masses. Namely, an SM singlet field is added in the bulk, which is then coupled to the Higgs and lepton doublet fields on IR brane.
%, giving the neutrino modes a Dirac-type mass from Higgs VEV (just like for SM-charged fermions).
%
Also, a Planck-size Majorana mass term for the bulk singlet field is turned on at the UV brane. 
Adding a Majorana mass only on the UV is natural due to an extended bulk EW gauge symmetry
(in turn, invoked in order to satisfy EW precision test bounds) under which the singlet {\em is} charged and which is broken on the UV brane.

Such a framework has all the makings
%
%/ingredients/germs/tell-tale signs/seeds/``smells'' 
%
of type I high-scale seesaw; 
indeed the bottomline formula for the SM neutrino mass in this model,
\bea
m_{ \nu } & \propto & \frac{ v^2 }{ M_N^{ \rm UV } },
\label{mnu_sketch_typeI}
\eea
%
%being a hallmark/tell-tale sign of such a seesaw mechanism, 
seems to conform to the above expectation (here, $M_N^{ \rm UV }$ is the Majorana mass term 
for the singlet on the UV brane). 
This result was derived in the earlier literature using the basis of the ``usual'' zero and KK modes, i.e., ignoring the Majorana mass term on the Planck brane, whose
effects are 
take into account subsequently in the form of Majorana mass terms for zero and KK modes.
The SM neutrino mass arises from exchange of only would-be zero mode with super-large Majorana mass term $M_N^{ \rm UV }$ in denominator of the above formula, with numerator being Dirac mass induced by Higgs VEV, just like the usual 4D seesaw.
On the other hand, the KK modes 
%(starting at $\sim$ TeV) 
contribute negligibly (even though they also have very large Majorana mass terms).

In this paper, we focussed instead on the {\em mass} basis for the singlet neutrino modes
(as might be required for studies involving {\em on}-shell production of the singlet neutrino states) and analyzed in detail neutrino mass generation via a 5D calculation.
%
%but also disscussed
%the CFT picture for more intuition.
%
Such a  change of basis
% -- seemingly innocuous -- 
actually turns out to lead to a paradigm shift. Our results show that the {\em same} formula for the SM neutrino mass, i.e., Eq.~(\ref{mnu_sketch_typeI}), should be reinterpreted as
\bea
m_{\nu} & \propto & \left( \frac{v^2}{\rm{TeV}} \right) \left( \frac{\rm{TeV}}{ M_N^{ \rm{UV}}} \right)
\label{mnu_sketch_inverse}
\eea
%
%(again, it also involves the fermion profiles in an intricate way, but this dependence is not essential to the fundamental point being made here.) 
%
%relevant for 
%big picture 
%
Namely, it is the exchange of TeV mass singlet modes with unsuppressed coupling to Higgs which
dominantly contribute to the SM neutrino mass, as indicated by 1st factor above. The smallness of the SM neutrino mass follows from these singlet modes being {\em mostly} Dirac with a very small fraction of their mass being Majorana-natured (which accounts for the 2nd factor). What is remarkable is that these highly suppressed Majorana mass terms are themselves completely natural being  the result of an incarnation of a seesaw mechanism 
%(i.e., do not require any sort of tuning) -- as seen by the {\em ratio} of scales entering 2nd factor, 
albeit here it is for the Majorana mass term for TeV-scale singlet modes!
This picture realizes a {\emph{natural}} version of a scenario dubbed ``inverse'' seesaw in the literature.
% (in the context of {\em four}-dimensional models).
%
The type I high-scale seesaw was 
merely
%
%just 
%
a mirage.
%Just to reiterate then, type I high-scale seesaw nature of the SM neutrino mass formula ({\em and} KK basis calculation thereof) might be somewhat of an ``illusion/mirage'' after all, since in the mass basis for the singlet modes it has ingredients of/recipe for inverse seesaw, that too a natural version of it (i.e., it comes with an -- built-in/automatic -- understanding of/explanation for smallness of the Majorana mass terms for TeV-scale singlet modes). 

Importantly, our finding leads to a radical shift in the {\em phenomenology} of this scenario. Indeed, we realized that the physical source of a dominant part the SM neutrino mass -- which are the TeV-mass singlet states -- can potentially be directly probed at colliders. Similarly, leptogenesis may occur at the TeV temperatures from decays of these singlet modes. The attention is therefore on TeV-scale physics.~\footnote{We will detail these ideas in ongoing work \cite{future}.}

We also discussed, for the first time, the CFT interpretation of this warped seesaw model.
The new ingredient relative to the case of charged SM fermions
%widely studied in the literature 
is the Majorana mass for the external singlet field coupled to the CFT.
Taking this into account 
%and simply chugging along/following one's nose, 
we confirmed that one naturally ends up with the inverse seesaw mechanism.
% i.e., this CFT picture is (much) closer to the 5D mass basis.
%
%In fact, by following the standard AdS/CFT dictionary, the CFT estimates for the SM neutrino mass and the appropriately renormalized mass of the external singlet
%
%that we found in the 5D mass basis
%calculation) 
%
%agreed with the 5D calculation.
% (the latter mass corresponding to that a super-heavy ``special'' mode in the 5D picture).
%
The CFT picture also clarifies the universal dependence on $c_N$ in the neutrino mass formula (\ref{mnu5D}), %[which was, for simplicity, suppressed in Eq.~(\ref{mnu_sketch_typeI}) just above], 
whose origin was somewhat obscured in the KK basis.

%So far, the discussion had been of mostly theoretical interest: after all neutrino mass formula is unchanged.
% (as expected)!
%
%-- future work : 
%
%-- LHC/future collider signals: probe neutrino mass generation directly
%
%

\section*{Acknowledgements}

%
%\begin{eqnarray}
%\end{eqnarray}
%

%

We would like thank Csaba Csaki, Rabindra Mohapatra and Raman Sundrum for discussions. This work was supported in part by NSF Grant No.~PHY-1315155 and Maryland Center for Fundamental Physics. SH was also supported in part by a fellowship from The Kwanjeong Educational Foundation. LV acknowledges the MIUR-FIRB grant RBFR12H1MW and the ERC Advanced Grant no. 267985 (DaMeSyFla). KA~would like to thank the Aspen Center for Physics for hospitality during the completion of this work.

\appendix

\section{Details of the 5D mass basis calculation}
\label{tedious}

\subsection{The 5D Model and KK decomposition}
\label{sec:app_model}

Varying the full action $S$ in (\ref{eq:action_1}) with respect to $\bar{\chi}$ and $\psi$ we get:
\bea 
-i \bar{{\sigma}}^\mu{\partial}_\mu \chi - \partial_5 \bar{\psi} + \frac{c_N+2}{z} \bar{\psi} = 0 \label{eq:EOM_1}\\
-i \sigma^\mu{\partial}_\mu \bar{\psi} + \partial_5 \chi + \frac{c_N-2}{z} \chi + d\frac{R}{z}\delta(z-R) \psi = 0 \label{eq:EOM_2}.
\eea
The boundary conditions in the absence of $S_{\rm UV}$ are chosen to be Dirichlet for $\chi$ (and consequently Neumann for $\psi$). The UV-Majorana mass alters the boundary conditions at $z=R$.

Following~\cite{Csaki:2003sh}, we slightly displace the UV-localized mass to $z=R+\epsilon$ and impose standard Dirichlet boundary conditions for $\chi$ at $z=R$. The effect of the localized mass is then encoded in a jump of the field: $\chi \vert_{R+\epsilon} = - d~ \psi \vert_{R+\epsilon}$. We can now send $\epsilon\to0$. The corresponding jump in $\psi$ may be found imposing the bulk equations of motion: $ \partial_5 \psi \vert_{R+\epsilon} = i d \slashed{\partial} \bar{\psi} \vert_{R+\epsilon}$.

Overall, the boundary conditions turn out to be: 
\bea
\chi \vert_{{R'}^-} = 0, ~~~~~~~~~~~~~  \chi \vert_{R^+} = - d~ \psi \vert_{R^+} \label{eq:BC_odd}.
\eea
For the sake of completeness, we also observe that the remaining two (redundant conditions) are $\partial_5 \psi \vert_{{R'}^-} = 0$, $\partial_5 \psi \vert_{R^+} = i d \slashed{\partial} \bar{\psi} \vert_{R^+}$.

Next, we perform a Kaluza-Klein reduction. Because the UV-localized mass breaks the $U(1)_\Psi$ number, the reduced 4D theory will be a dynamics of Majorana fermions. It is therefore convenient to decompose $\chi,\psi$ in terms of a single tower of Weyl fermions:
\beq \label{eq:KK_decomp}
\chi (x, z) = \sum_n g_n(z) \xi_n(x), ~~~~~~~~~~~~ \bar{\psi} (x, z) = \sum_n f_n(z) \bar{\xi}_n (x),
\eeq
where $\xi_n$ satisfy Majorana equations of motion $-i \bar\sigma^\mu{\partial}_\mu \xi_n + m_n \bar{\xi}_n = 0$. 
The bulk equations of motion and the boundary conditions then become
\ba\label{eq:EOM}
f_n' + m_n g_n - \frac{c_N+2}{z} f_n = 0~~~~~&&~~~~~ g_n' - m_n^* f_n + \frac{c_N-2}{z} g_n = 0, \label{eq:EOM_g_n_2_bulk}\\\no
g_n(R')=0~~~~~&&~~~~~g_n(R)=-df^*_n(R).
\ea
The Dirac mass parameter $c_N$ is real by Hermiticity of the action. In addition, by making a phase rotation of $\psi$ we can always eliminate the phase in $d$. Since $\psi$ is one component of $\Psi$, in order not to break 5D Lorentz invariance, we are actually performing a phase rotation of the 5D field $\Psi$ itself. We conventionally take $d>0$ from now on. Finally, $m_n$ are real because they are the eigenvalues of the Hermiticitan differential operator defined by Eqs~(\ref{eq:EOM}) in the metric determined by the kinetic term. Hermiticity also guarantees that the Kaluza-Klein expansion (\ref{eq:KK_decomp}) is meaningful.

Consistently, observe that inserting (\ref{eq:KK_decomp}) in (\ref{eq:action_1}) gives
\ba \label{eq:action_full_for_normalization_KK_simplified_final}
S =\int d^4x \left[\int dz \sum_{n,m}  \left( \frac{R}{z} \right)^4  \left( f_n^* f_m + g_n^* g_m \right) \right] \left\lbrace - i \xi_n \slashed{\partial} \bar{\xi_m} + \frac{1}{2} \left( m_n^* \xi_n \xi_m + m_n \bar{\xi_n} \bar{\xi_m} \right) \right\rbrace,
\ea
The normalization is therefore defined by 
\beq \label{eq:normalization_condtion_f_n_g_n}
\int dz \left( \frac{R}{z} \right)^4  \left( f_n^* f_m + g_n^* g_m \right) = \delta_{nm}.
\eeq

%We can justify the decomposition (\ref{eq:KK_decomp}) by defining a unified 4-component vector
%\ba
%\Phi= \left( \begin{array}{c}\phi_R \\ \phi_I \end{array} \right)
%\ea
%with the boundary condition following (\ref{eq:KK_decomp}) and the same metric. This is now a complete set.

For clarity we stress our convention for $c_N$, which we do by solving the zero mode equation for the right-handed fermion $g_n$, i.e. Eq(\ref{eq:EOM}) with $m_n = 0$. By plugging the solution into the action, one can easily see that $c_N = -1/2$ (as opposed to 1/2) corresponds to flat, $c_N > -1/2$ a IR-localized and $c_N < -1/2$ a UV-localized profile.

We decide to carry out the Kaluza-Klein decomposition with real eigenfunctions $f_n, g_n$ (as in \cite{Vecchi:2013xra}), in which $m_n$ are allowed to acquire any (real) positive or negative value.~\footnote{One may alternatively work with both real and imaginary components of the wave-functions, but with a constraint $m_n>0$ on the eigenvalues (we believe this is the convention implicitly adopted in \cite{Huber:2003sf}). We checked that our results do not depend on which convention is used.}} Before proceeding with the actual calculation of the spectrum, note that the eigenvalue problem is invariant under the following spurious symmetry:
\ba\label{spurious}
(f_n, g_n,m_n,d)\to(f_n,-g_n,-m_n,-d).
\ea
This tells us that for $d=0$ the solution consists of Dirac pairs: there exists an {\emph{independent}} solution with eigenvalue $-m_n$ for any eigenfunction with mass $m_n$. This is no more true as soon as $d\neq0$, and no exact pairing is observed.

%In the general case with non-vanishing $d$, we identify the following symmetry (triviality...): $$ (f_n, g_n,m_n,d)\to(\pm if_n,\mp ig_n,-m_n,d),$$ This is a triviality in terms of the equation $\phi\to(\Delta\phi)^*$ where $\phi^t=(g,\bar g)$ because: $$(f_n, \bar g_n,m_n,d)\to(\pm if_n,\pm i\bar g_n,-m_n,d),$$ which describes an exchange between real and imaginary components (up to a sign) for a given $d$. This is saying that (for arbitrary $d$) the solution with eigenvalue $-m_n$ may be obtained from the eigenfunction with $m_n$ by exchanging real and imaginary parts (up to a sign). In other words,  the spectrum may alternatively be determined fixing $m_n>0$, provided we retain both imaginary and real parts of $f_n, g_n$.

The coupled system described by the bulk equations of motion can be decoupled in a straightforward way, yielding Bessel equation. The result is given by:
 \ba\label{sol}
g_n(z) &=& -\frac{1}{N_n}\frac{m_n}{\vert m_n \vert} z^{5/2}\left[ J_{-c_N-1/2} (\vert m_n \vert z) +b_n Y_{-c_N-1/2} (\vert m_n \vert z)\right]\\\no
f_n (z) &=& \frac{1}{N_n}z^{5/2}\left[J_{-c_N+1/2} (\vert m_n \vert z) +b_n Y_{-c_N+1/2} (\vert m_n \vert z) \right].
\ea
The coefficient $b_n$ is constrained by the boundary conditions:~\footnote{%An analogous expression was used in \cite{Contino:2004vy}:
This is equivalent to the alternative solution:
\ba\label{SH}
g_n(z) &=& \frac{m_n}{\vert m_n \vert} z^{5/2} \left[ C_n J_{c_N+1/2} (\vert m_n \vert z) - D_n J_{-c_N-1/2} (\vert m_n \vert z)\right]\\\no
f_n (z) &=& z^{5/2} \left[ C_n J_{c_N-1/2} (\vert m_n \vert z) + D_n J_{-c_N+1/2} (\vert m_n \vert z) \right].
\ea
Indeed, using $Y_{\nu} (x) = \frac{J_{\nu}(x) \cos (\nu \pi) - J_{-\nu} (x)}{\sin (\nu \pi)}$ we get:
\ba
C_n=-\frac{1}{N_n}\frac{b_n}{\cos(c_N\pi)}~~~~~~~~~~~D_n=\frac{1}{N_n}\left(1+{b_n}{\tan(c_N\pi)}\right).
\ea
In particular, $D_n/C_n=-\cos(c_N\pi)/b_n+\sin(c_N\pi)$. The authors independently checked all results of the paper using both (\ref{SH}) and (\ref{sol}).
}
%%%%%%%%%%%%
\ba\label{eq:BC_g_n}
-b_n=\frac{J_{-c_N-1/2}(|x_n|)}{Y_{-c_N-1/2}(|x_n|)}=\frac{J_{-c_N-1/2}(|x_n|/\Omega)-d\frac{x_n}{|x_n|}J_{-c_N+1/2}(|x_n|/\Omega)}{Y_{-c_N-1/2}(|x_n|/\Omega)-d\frac{x_n}{|x_n|}Y_{-c_N+1/2}(|x_n|/\Omega)},
\ea
where $x_n=m_nR'$ and $\Omega \equiv{R'}/{R}$. This is the equation constraining the eigenvalues $x_n$. Defining $Z_\nu(y)\equiv J_\nu(y)+b_nY_\nu(y)$, the normalization is determined by
\ba\label{norm}
N_n^2&=&R^4\int^{R'}_R dz~z\left[Z_\nu^2(|m_n|z)+Z_{\nu+1}^2(|m_n|z)\right]\\\no
&=&\frac{R^4}{2}\left({\cal I}_n(R')-{\cal I}_n(R)\right),
\ea
where $\nu=-c_N-1/2$ and ${\cal I}_n(z)=z^2\left[Z_\nu^2(y)-Z_{\nu+1}(y)Z_{\nu-1}(y)+Z_{\nu+1}^2(y)-Z_{\nu+2}(y)Z_{\nu}(y)\right]$, $y=|m_n|z$.

\subsection{Masses}
\label{sec:app_masses}

We can find approximate analytic solutions for the modes satisfying $|x_n|\ll\Omega$. %; Transplanckian states with $|x_n|\gtrsim\Omega$ are not relevant to the calculation of the Standard Model neutrino mass. \sh{$\leftarrow$ needs more clarification why ? coupling is $\mathcal{O}(1)$ but too large mass ?} 
Using a small argument approximation of the Bessel functions for the UV boundary condition, the spectrum equation (\ref{eq:BC_g_n}) is simplified to
\begin{adjustwidth}{-1.2cm}{}
\ba \label{eq:b}
-{b_n}&=&\frac{J_{-c_N-1/2}(|x_n|)}{Y_{-c_N-1/2}(|x_n|)}%\approx\frac{1}{\tan \left(|x_n|+ \frac{c}{2} \pi \right)} 
\approx\frac{1}{\frac{\Gamma^2(-c_N+1/2)}{\pi} \left(\frac{|x_n|}{2 \Omega} \right)^{2c_N} \left[ d\frac{x_n}{|x_n|} + \frac{1}{(c_N+1/2)} \left( \frac{|x_n|}{2 \Omega} \right) \right] + \tan (c_N \pi)}.
\ea
\end{adjustwidth}
To derive this expression we assumed $c_N\neq-1/2$. From now onwards we will consider $c_N<0$. We will also assume that $d$ is smaller than one, but much larger than the TeV-Planck hierarchy. 

The ratio $b_n$ can also be approximated for large arguments $|x_n|\gg1$ by $b_n\approx\frac{1}{\tan \left(|x_n|+ \frac{c_N}{2} \pi \right)}$. However, this approximation will break down for the first few KK modes. Because, as we will show below, these give the most important contribution to the SM neutrino mass, we keep the general expression (\ref{eq:b}) for now.

For $c_N<0$ and $|{x_n}|/\Omega\ll d$ (and far from special points discussed shortly), $\tan(c_N\pi)$ can be neglected from the right-hand side of Eq.~(\ref{eq:b}) and
\ba
-b_n=\frac{J_{-c_N-1/2}(|x_n|)}{Y_{-c_N-1/2}(|x_n|)}%\approx\frac{1}{\tan \left(|x_n|+ \frac{c}{2} \pi \right)}
\approx\frac{x_n}{|x_n|}\frac{\pi}{d\Gamma^2(-c_N+1/2)}\left( \frac{|x_n|}{2 \Omega} \right)^{-2c_N}.
\ea
As can be seen from $|b_n|\propto\left({|x_n|}/{\Omega} \right)^{-2c_N}\ll1$, the spectrum of light modes is approximately determined by $x_n=\pm x_n^0$, where $x_n^0$ are the zeros of $J_{-c_N-1/2}$. For $n$ not too small, using the large argument expansion, these are approximately given by $x_n^0\approx\left( n + \frac{1}{2}(1-c_N) \right) \pi$ with $n=0,1,\cdots$. Including the leading correction we get 
\ba\label{massLight}
x_n&=&\pm x_n^{0}+\delta_n\\\no
\delta_n&=&\frac{Y_{-c_N-1/2}(|x_n^0|)}{J'_{-c_N-1/2}(|x_n^0|)}\,\frac{\pi}{d\Gamma^2(-c_N+1/2)}\left( \frac{|x_n^{0}|}{2 \Omega} \right)^{-2c_N}.
\ea
This result shows that the light modes are approximately Dirac pairs~\footnote{A Dirac fermion consists of two Weyl fermions of mass $\pm m$.} up to a split $\delta_n$, induced when the UV-localized Majorana mass is turned on. In other words, there are two towers of Weyl spinors, one with positive masses (``positive tower'') and the other with negative masses (``negative tower''); the modes with $|{x_n}|/\Omega\ll d$ (``low-lying modes'') form pseudo-Dirac pairs.

In the vicinity of the zeros of the denominator of the right-hand side of (\ref{eq:b}), the function $b_n$ is no more much smaller than one and we need a separate analysis. In this regime the mass eigenstates are identified by the fact that the denominator of the right-hand side of (\ref{eq:b}) is much smaller than one (or very close to zero):
%this equation:
\ba \label{eq:eq_for_special_mode_c_less_1/2}
\frac{d}{\pi} \Gamma^2(-c_N+1/2)\left(\frac{|x_n^{\text{special}}|}{2 \Omega} \right)^{2c_N} \left[ \frac{x^{\text{special}}_n}{|x^{\text{special}}_n|} + \frac{1}{d (c_N+1/2)} \left( \frac{|x^{\text{special}}_n|}{2 \Omega} \right) \right] + \tan (c_N \pi)\approx 0.
\ea
%\sh{Again, following the comments from ``draft'', here I will distinguish ``single-special'' vs. ``paired-special''. So, I am rewriting most of paragraphs below, trying to clarify/supplement some points. Of course, we can change if we have better/clearer option.}
As we will see shortly, the mode $x^{\text{special}}_n$ that satisfies (\ref{eq:eq_for_special_mode_c_less_1/2}) is {\emph{special}} in the sense that there is no analog solution of mass $\sim-x^{\text{special}}_n$, that is, it is \emph{un}paired (and so pure Majorana), unlike the usual cases where there are two modes in each TeV-bin, making up a (pseudo) Dirac pair. For this reason, we will call such mode ``\emph{single-special}'' mode. Later, we will introduce ``\emph{paired-special}'' modes, which, as the name indicates, consist of a pair of two Weyl fermions of mass close to the single-special and a mass splitting of order the TeV.

%As we will soon show, the special modes are at $x_n>0$ in our convention $d>0$.  

Now, let us discuss in detail when (\ref{eq:eq_for_special_mode_c_less_1/2}) can be satisfied. Consider first $-1/2<c_N<0$, for which $\tan(c_N\pi)<0$. If $|x^{\text{special}}_n|\gtrsim d\Omega$ the second term in the squared parenthesis dominates over the first term. In this case %, whichever value of $x_n$ satisfies the equation it is $d$-independent.\footnote{It will be rather strange if there exists a pure Majorana mode which is completely independent of Majorana mass ($d$) on UV brane, the only source of lepton number violation.} And, perhaps more importantly, 
since $2 c_N + 1 > 0$, $\left( |x_n|/\Omega \right)^{2 c_N + 1} \ll 1$ for $\forall |x_n| \ll \Omega$ and yet, for generic value of $c_N \in (-1/2,0)$, $\tan (c_N \pi) \sim \mathcal{O}(1)$. That is, for a generic value of $c_N$ (\ref{eq:eq_for_special_mode_c_less_1/2}) cannot be satisfied by modes below $\Omega$. On the other hand, when $|x^{\text{special}}_n|\ll d\Omega$ the first term in the squared parenthesis dominates. Because $\tan(c_N\pi)<0$, the cancellation can occur only when the first term is positive, i.e. the solution exists only for $x^{\text{special}}_n>0$. The solution is given by:
\ba\label{special>}
\frac{x_n^{\text{special}}}{2 \Omega} \approx \left(\frac{-\pi \tan (c_N \pi)}{d\Gamma (-c_N+1/2)^2}\right)^{\frac{1}{2c_N}},~~~~~~~~~~-\frac{1}{2}<c_N<0.
\ea
We stress out again that $x_n^{\text{special}}\ll d\Omega$ and, as anticipated, there is no analog behavior at $x_n<0$. This is how we see that the ``single-special'' mode is unpaired.

For $c_N\lesssim-1/2$, the second term of (\ref{eq:eq_for_special_mode_c_less_1/2}) is negative and $\tan(c_N\pi)>0$. Again, when $|x^{\text{special}}_n|\gtrsim d\Omega$ the second term in the squared parenthesis dominates. However, as in the previous case, no solution is found when $d\Omega\lesssim|x^{\text{special}}_n|<\Omega$ for generic choice of $c_N < -1/2$. Similarly, for $|x^{\text{special}}_n|\ll d\Omega$ the first term dominates and one would seem to find $|x_n|\sim\Omega d^{-\frac{1}{2c_N}}$; however, this value is now much larger than $d$, and is therefore inconsistent with the original hypothesis $|x^{\text{special}}_n|\ll d\Omega$. A solution is only possible when the terms inside the squared parenthesis approximately cancel each other. This is possible only when $x_n >0$ and thus mass of the special mode is in the positive tower (i.e. $x_n>0$) and parametrically close to the UV-localized Majorana mass:
\beq \label{eq:mass_special_mode_c_bigger_1/2}
\frac{x_n^{\text{special}}}{2 \Omega} \sim - (c_N+1/2) d,~~~~~~~~~~c_N<-\frac{1}{2}.
\eeq
Again, no partner at $-x_n^{\text{special}}$.

In summary, with our convention $d>0$ the single-special mode is located in the positive tower for both $c_N >$ or $< -1/2$ albeit with parametrically different mass for single-special mode. No special behavior (i.e. no singularity in the right-hand side of (\ref{eq:b})) is present in the negative tower.

We conclude this section with a few more comments on the spectrum. We start with $-1/2<c_N<0$. In this case, since $|x^{\rm special}|\ll d\Omega$, the analysis leading to (\ref{massLight}) allows us to conclude that all states with mass $|x_n|\ll|x^{\rm special}|$ are pseudo-Dirac with mass splitting of order $\delta_n$. The denominator of (\ref{eq:b}) gets smaller as we approach the special mode in the positive tower, whereas $b_n$ remains very small for $x_n\sim-|x^{\rm special}|$. This suffices to argue that the mass splitting for states close to the special mode is generically of order the TeV ($\delta_n\sim1$). These pseudo-Dirac fermions have mass splitting (of order the TeV) much smaller than their mass $\sim|x^{\rm special}|$ but much larger than that of low-lying modes. %The origin of such a large mass splitting is of course the fact that, unlike low-lying modes, the denominator of the right-hand side of (\ref{eq:b}) is decreased to $\lesssim \mathcal{O} (1)$ in one of the tower (positive tower in our case), and this was what makes single-special mode ``special''. 
We call them ``paired-special'' modes. 
% Therefore, unless strange parametric tuning happens such that the positive tower passes through zero very close to the zeros of $J_{-c_N-1/2}(|x_n|)$ (around which the mass of the negative tower arises), generically the size of mass splitting $\delta_n$ will be of ${\cal O}(1)$ when $|x_n|\sim x^{\rm special}$ (we call these modes ``paired-special'' modes as explained below for $c_N<-1/2$). 

The states heavier than the special mode are again pseudo-Dirac, with a mass splitting controlled by $|b_n|\ll1$ between $x_n^{\rm special}\ll|x_n|\ll d\Omega$.

When $c_N<-1/2$ the states with $|x_{n}|\ll d\Omega$ are pseudo-Dirac with mass splitting $\delta_n$. However, since $x_n^{\rm special}\sim d\Omega$ our equation (\ref{massLight}) breaks down before we reach the special mode; to precisely estimate the mass splitting for $|x^{\rm special}|\lesssim d\Omega$ one may perform a completely analogous analysis without dropping $\tan (c_N \pi)$. We do not quote the result for brevity. The modes at $x_n^{\rm special}\sim d\Omega$ have $b_n={\cal O}(1)$ and typically a Majorana splitting of order the TeV, which is the maximal value set by the IR brane. As above, for $|x_n|\gtrsim d\Omega$ the states are pseudo-Dirac.

%It is interesting to ask how many such paired-special modes exist, that is what is the width of the special point (\ref{eq:mass_special_mode_c_bigger_1/2}). 

As we will discuss below, in order to make sense of the SM neutrino mass calculation in the case of $c_N < -1/2$ it is useful to know the number of the paired-special modes. We can address this question by determining the width of the special point (\ref{eq:mass_special_mode_c_bigger_1/2}), i.e. what condition on $\eta=x_n-x_n^{\rm special}$ follows requiring the right-hand side of (\ref{eq:eq_for_special_mode_c_less_1/2}) is allowed to be of order unity (or more precisely, of $\mathcal{O} (\tan (c_N \pi))$). This gives:
\ba
\eta\lesssim\tan(c_N\pi)\frac{2\pi(-1/2-c_N)^{1-2c_N}}{\Gamma^2(-c_N+1/2)}\Omega d^{-2c_N}.
\ea
With realistic numbers (say, $c_N=-0.7$, $d=10^{-3}$, $\Omega\sim 10^{15}$), one finds $\eta\gg1$ ($5\times10^{8}$). %One may wonder why we didn't worry about this issue for $-1/2 < c_N < 0$. A complete answer will be discussed shortly, but just to give a short answer, in the case of $-1/2 < c_N < 0$, the contribution from modes with mass $\gg \rm TeV$ to the SM neutrino mass is negligible. On contrary, for $c_N < -1/2$, the contribution from the single/paired-special modes are overwhelmingly large and it is crucial to carefully estimate the size of such contributions in order to explain the final ``right'' answer. 

%{\color{blue}Finally, the heavier states $|x_n|\gg x^{\rm special}$ are again pseudo-Dirac pairs with a small splitting of order $b_n\sim(|x_n|/\Omega)^{-(2c_N+1)}\ll1$. This can be easily seen from Eq(\ref{eq:b}) by noting that for $|x_n|\gg x^{\rm special} \sim d \Omega$ it is the second term in the squared parenthesis that dominates and it is in fact much larger than one and hence one can drop $\tan (c_N \pi)$. Notice that the mass splitting for modes above the special modes are (parametrically) smaller than that of low-lying modes by a factor of $\sim \frac{d \Omega}{\vert x_n \vert} \ll 1$, and this factor is crucial for the convergence of sum of contributions from these modes to the SM neutrino mass.  }

%We numerically verified that our analytical expressions (\ref{special>}) and (\ref{eq:mass_special_mode_c_bigger_1/2}) are accurate. 

\subsection{Couplings}
\label{sec:app_couplings}

We are interested in the couplings of $\xi_n$ to the zero mode $L(x)$ of $\Psi_L$, that we identify with the Standard Model lepton doublet: 
%\sh{The convention for $c_L$ is $c_L = 1/2$ (not $-1/2$) is flat ? Is is okay to use different conventions for $c_N$ and $c_L$ ? I mean I know it is okay technically, but I am asking it for presentation purpose..}
\ba
\Psi_L\to \Psi_L^{(0)}(z)L,~~~~~~~~~~~\Psi_L^{(0)}=\frac{1}{\sqrt{R}}\sqrt{\frac{2c_L-1}{1-\Omega^{1-2c_L}}} \left( \frac{z}{R} \right)^{2-c_L},
\ea
where $M_L=c_L/R$ is the 5D mass of $\Psi_L$. Introducing the canonically normalized 4D field $H=R'/R{\cal H}$, eq(\ref{eq:4D_Yukawa_eff_1}) becomes:
\ba\label{eq:4D_Yukawa_eff_1'}
\delta S &=& -\int d^4x~y_nHL\bar\xi_n,
\ea
where
\ba\label{eq:y_n}
y_n&=&\Omega^{-3}\lambda_5\Psi^{(0)}_L(R')f_n(R').
%\\\no&=&\Omega^{-1-c_L}\frac{\lambda_5}{\sqrt{R}}\sqrt{\frac{2c_L-1}{1-\Omega^{1-2c_L}}} ~f_n(R')\\\no
%&=&\Omega^{-1-c_L}\frac{\lambda_5}{\sqrt{R}}\sqrt{\frac{2c_L-1}{1-\Omega^{1-2c_L}}}R'^{5/2}\frac{1}{N_n}Z_{\nu+1}(|m_n|R').
\ea
The wave function $\Psi_L^{(0)}(R')$ can be read from above. The profile of the singlet can be written as $f_n(R')=R'^{5/2}Z_{\nu+1}(|m_n|R')/N_n$, where $Z_\nu=J_\nu+b_nY_\nu$ with $\nu=-c_N-1/2$. We will now carefully determine $f_n(R')$ for the low-lying (pseudo-Dirac) modes $|x_n|\ll x_n^{\rm special}$. The coupling for modes around $x_n^{\rm special}$ will be analyzed subsequently.

%The large argument limit of the Bessel function can be used to show that
%\ba\label{ZnuIR}
%Z_{\nu\pm1}&\to&\pm\sqrt{\frac{2}{\pi|x_n|}}(-1)^n+{\cal O}(\delta^2)
%\\\no&=&\pm\sqrt{\frac{2}{\pi|x_n|}}(-1)^n+{\cal O}(\delta^2)
%\ea
The normalization (\ref{norm}) receives a contribution from $z=R'$ and one from $z=R$. To analyze the former we observe that the boundary condition for $g_n(z)$ in the IR implies $Z_\nu(|m_n|R')=0$ (see Eq(\ref{eq:EOM})). Then, from the definition (\ref{norm}), and using the identity $Z_{\nu+1}(|x_n|)+Z_{\nu-1}(|x_n|)=\frac{2\nu}{|x_n|}Z_\nu(|x_n|)=0$, we get ${\cal I}_n(R')=R'^2[-Z_{\nu+1}Z_{\nu-1}+Z_{\nu+1}^2](|x_n|)=2R'^2Z_{\nu+1}^2(|x_n|)$. 

 In the UV the boundary condition reads $Z_\nu(|x_n|/\Omega)=d~(x_n/|x_n|)Z_{\nu+1}(|x_n|/\Omega)$. We are interested in ${\cal I}_n(R)$, the UV contribution to the normalization $N_n$. For $|x_n|\ll|x_n^{\rm special}|$ we can use the small argument approximation of the Bessel functions. At leading order, when $c_N\neq-1/2$ (and $c_N<1/2$), the relevant expressions are:
\ba\label{eq:ZZZ}
 Z_\nu(|x_n|/\Omega)&\sim&\left(\frac{|x_n|}{2\Omega}\right)^\nu\frac{1}{\Gamma(\nu+1)}\left[1+{\cal O}(\delta, |x_n|/\Omega)\right],\\\no
 Z_{\nu-1}(|x_n|/\Omega)&\sim&\left(\frac{|x_n|}{2\Omega}\right)^{\nu-1}\frac{1}{\Gamma(\nu)}\left[1+{\cal O}(\delta,( |x_n|/\Omega)^3)\right],\\\no
 Z_{\nu+2}(|x_n|/\Omega)&\sim&-b_n\left(\frac{2\Omega}{|x_n|}\right)^{\nu+2}\frac{\Gamma(\nu+2)}{\pi}\left[1+{\cal O}(|x_n|/\Omega)^3\right].
 \ea
 In order to understand whether the subleading ${\cal O}(\delta, |x_n|/\Omega)$ terms must be kept in our analysis we have to compare the leading order estimate of ${\cal I}_n(R)$ with ${\cal I}_n(R')\sim R'^2/|x_n|$. The leading contribution of $Z_\nu^2$ and $Z_{\nu+1}^2$ to ${\cal I}_n(R)$ are suppressed by $|x_n|/\Omega$ compared to the other two and can be neglected. The dominant terms give ${\cal I}_n(R)\sim R^2(|x_n|/\Omega)^{2\nu-1}\sim R^2\delta_n(|x_n|/\Omega)^{-2}\sim R'^2\delta_n/|x_n|^2$, which is itself a correction of order $\delta_n/|x_n|$ of $N_n$. Being interested in corrections at most of order $\delta$ in the normalisation $N_n$, we can safely neglect $O(\delta)$ terms in (\ref{eq:ZZZ}), since they lead to ${\cal O}(\delta_n^2)$ corrections in $N_n$. A more accurate calculation gives
\ba
{\cal I}_n(R)R^{-2}&=&\left(-\frac{x_n}{|x_n|}\frac{1}{d}Z_\nu Z_{\nu-1}-Z_\nu Z_{\nu+2}\right)\left[1+{\cal O}(|x_n|/\Omega)\right]\\\no
%&\approx&-\frac{x_n}{|x_n|}\left(\frac{|x_n|}{2\Omega}\right)^{2\nu-1}\frac{\nu}{d\Gamma^2(\nu+1)}-\frac{x_n}{|x_n|}\frac{\nu+1}{d\Gamma^2(\nu+1)}\left(\frac{|x_n|}{2\Omega}\right)^{2\nu-1}\\\no
&=&-\frac{x_n}{|x_n|}\left(\frac{|x_n|}{2\Omega}\right)^{2\nu-1}\frac{2\nu+1}{d\Gamma^2(\nu+1)}\left[1+{\cal O}(\delta, |x_n|/\Omega)\right]\\\no
&=&-\frac{x^0_n}{|x^0_n|}\delta_n\frac{J_\nu'(|x_n^0|)}{Y_\nu(|x_n^0|)}\left(\frac{|x^0_n|}{2\Omega}\right)^{-2}\frac{2\nu+1}{\pi}\left[1+{\cal O}(\delta)\right].
%\\\no&\to&\frac{x^0_n}{|x^0_n|}\delta_n\frac{2\nu+1}{\pi}\left(\frac{|x^0_n|}{2\Omega}\right)^{-2}\left[1+{\cal O}(\delta)\right].
\ea
In the second step we replaced (\ref{eq:ZZZ}) and used the definition of $b_n$ given in (\ref{eq:b}). In the third step we neglected the correction arising from the replacement $x_n\to x_n^0$, since in our final estimate of $N_n$ it would appear as a ${\cal O}(\delta^2)$ effect, which we drop.

Summing the UV and IR contributions we find
 \ba
 N^2_n&=&R^4R'^2Z_{\nu+1}^2(|x_n|)\left[1-2\frac{x^0_n}{|x^0_n|}c_N\frac{\delta_n}{|x^0_n|}\left(\frac{2}{\pi|x^0_n|}\frac{J_\nu'(|x^0_n|)/Y_\nu(|x^0_n|)}{Z_{\nu+1}^2(|x^0_n|)}\right)\right].
 %\\\no&\to&\frac{2R^4R'^2}{\pi|x_n|}\left[1+2\frac{x_n}{|x_n|}c\frac{\delta_n}{|x_n|}\right]
 \ea
For later convenience we factored out $Z_{\nu+1}^2(|x_n|)$ because it automatically cancels out in the expression $f_n/N_n$ entering $y_n$. This results in a $1/Z_{\nu+1}^2(|x_n|)$ factor in the $\delta_n$ correction. Despite the fact that $|x_n|=|x^0_n|(1+x^0\delta_n/|x^0_n|^2+\cdots)$, Because we content ourselves with ${\cal O}(\delta_n)$ effects, we can safely replace $x_n\to x^0_n$ in the squared parenthesis. On the other hand, the overall $Z_{\nu+1}^2(|x_n|)$ contributes an additional ${\cal O}(\delta_n)$ term to $N_n$, but --- as anticipated --- this effect cancels out from (\ref{eq:y_n}). More precisely, putting everything together we get: 

\begin{adjustwidth}{-1.2cm}{}
\ba\label{eq:y}
y_n%&=&\Omega^{-1-c_L}\frac{\lambda_5}{\sqrt{R}}\sqrt{\frac{2c_L-1}{1-\Omega^{1-2c_L}}}R'^{5/2}\sqrt{\frac{\pi|x_n|}{2R^4R'^2}}\sqrt{\frac{2}{\pi|x_n|}}(-1)^n\left[1+\frac{x_n}{|x_n|}c\frac{\delta_n}{|x_n|}\right]\\\no
&=&\frac{\lambda_5}{R}\sqrt{\frac{2c_L-1}{1-\Omega^{1-2c_L}}}\Omega^{1/2-c_L}{\rm sign}(Z_{\nu+1})\left[1+\frac{x_n}{|x_n|}c_N\frac{\delta_n}{|x_n|}\left(\frac{2}{\pi|x^0_n|}\frac{1}{J_{\nu+1}^2(|x_n^0|)}\frac{J'_\nu(|x_n^0|)}{Y_\nu(|x_n^0|)}\right)\right].
%\\&\to&\frac{\lambda_5}{R}\sqrt{\frac{2c_L-1}{1-\Omega^{1-2c_L}}}\Omega^{1/2-c_L}(-1)^n\left[1-\frac{x_n}{|x_n|}c\frac{\delta_n}{|x_n|}\right],
\ea
\end{adjustwidth}
This result holds for $|x_n|\ll x_n^{\rm special}$ up to terms of order $\delta_n^2$.

We now turn to a discussion of the couplings of the modes of mass near $x_n^{\rm special}$, which correspond to the special mode and the paired spacial modes. States in the negative tower always have $|b_n|\ll1$ and may be analyzed in a way completely analogous to what we have done for the light modes. The result is:
\ba\label{eq:yspecial-}
y_n&=&\frac{\lambda_5}{R}\sqrt{\frac{2c_L-1}{1-\Omega^{1-2c_L}}}\Omega^{1/2-c_L}{\rm sign}(Z_{\nu+1})\left[1+{\cal O}(b_n)\right].
\ea
In the positive tower the crucial difference is that $b_n$ is unsuppressed. This implies that our estimate of the UV contribution to the normalization $N_n$ must take this into account. In particular, (\ref{eq:ZZZ}) are no more accurate. Instead, assuming $b_n={\cal O}(1)$ we find that ${\cal I}_n(R)\sim R^2Z_\nu Z_{\nu+2}\sim R^2(|x_n|/\Omega)^{-2\nu-2}\sim {\cal I}_n(R')|x_n|^{2c_N}\Omega^{-2c_N-1}$. The subleading terms are of order $(|x_n|/\Omega)^{-2c_N}$ and $(|x_n|/\Omega)$. Neglecting them, we conclude that
\ba\label{eq:yspecial+}
y_n&=&\frac{\lambda_5}{R}\sqrt{\frac{2c_L-1}{1-\Omega^{1-2c_L}}}\Omega^{1/2-c_L}{\rm sign}(Z_{\nu+1})\left[1+a|x_n|^{2c_N}\Omega^{-2c_N-1}\right],
\ea
where $a$ is some number of order one. Finally, for the special mode it is not possible to determine $b_n$ analytically (it may well be that $|b_n|\gg1$, so the previous derivation does not apply). Yet, for any $b_n$ we expect
\ba
y_n^{\rm special}\sim\frac{\lambda_5}{R}\sqrt{\frac{2c_L-1}{1-\Omega^{1-2c_L}}}\Omega^{1/2-c_L}.
\ea
This estimate is correct up to a number of order unity.

\subsection{SM neutrino mass for $-1/2<c_N<0$}
\label{sec:app_SM neutrino mass for c_N > -1/2}

%%%%%%%%%%%%%%%%%%%
\begin{figure}
\begin{center}
\centering
\mbox{\includegraphics[height=50mm]{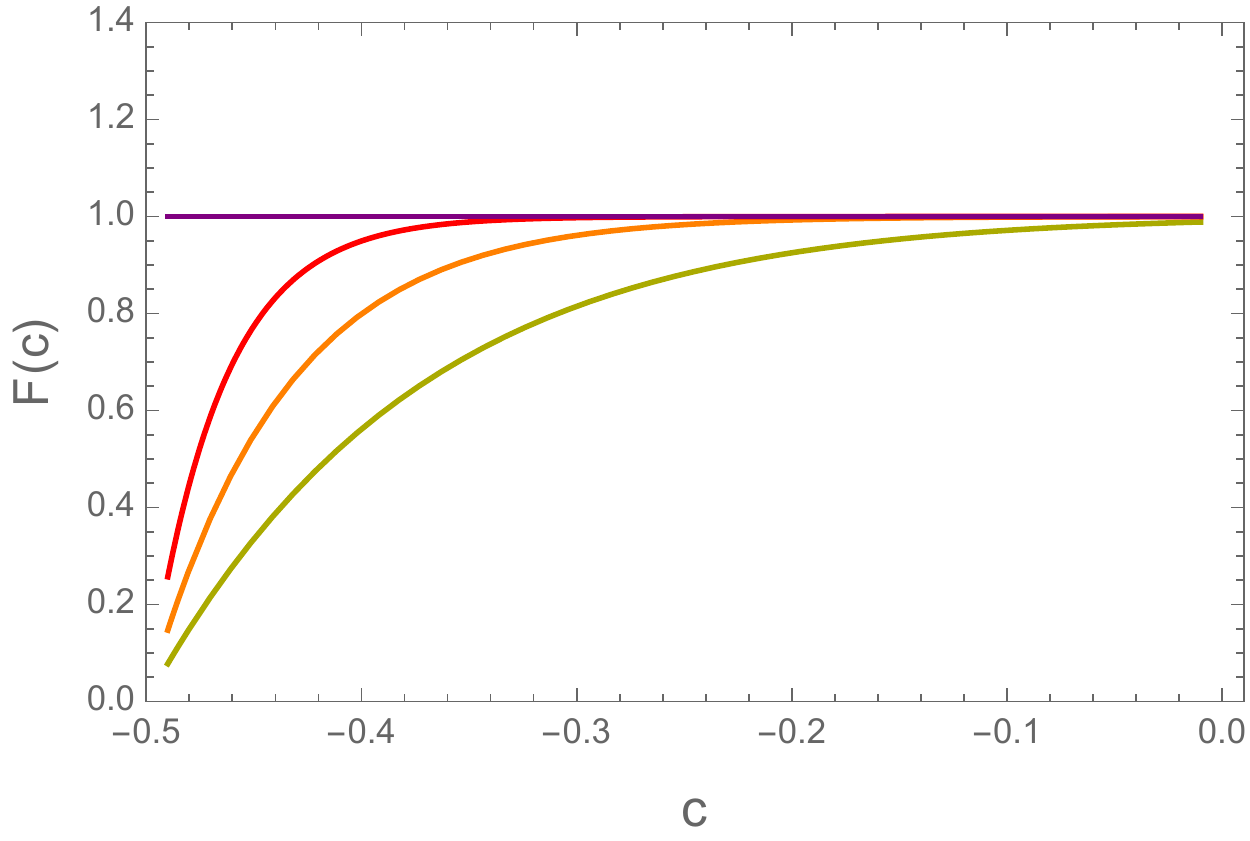}}~~~~\mbox{\includegraphics[height=40mm]{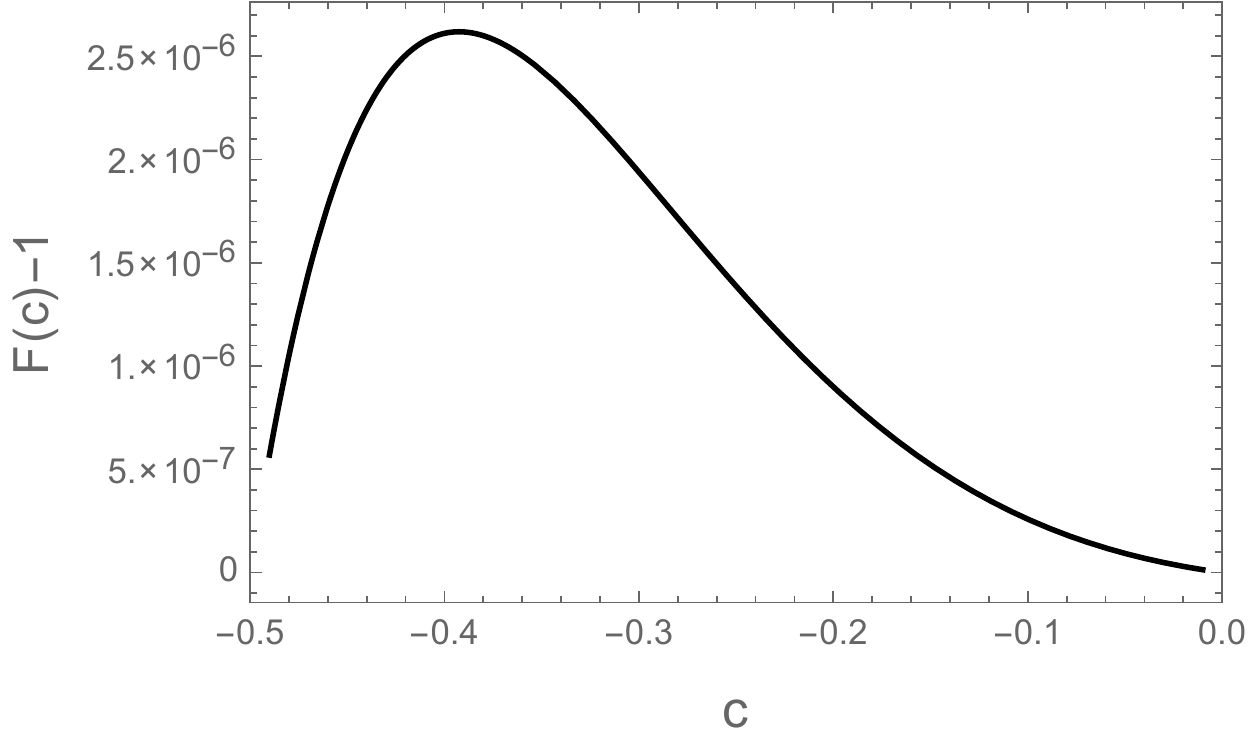}}
\caption{Function $F(c_N)$ defined in (\ref{F}) for $n_{\rm max}=10$ (yellow), $10^3$ (orange), $10^6$ (red), $\infty$ (purple). For comparison we also show the deviation of the purple line from $1$ (right plot).}
\label{fig:agree}
\end{center}
\end{figure}
%%%%%%%%%%%%%%%%%

The relevant part of the Lagrangian is (must change notation):
\ba
{\cal L}=\frac{m_n}{2}\bar\xi_n\bar\xi_n-y_nHL\bar\xi_n+{\rm hc}.
\ea
Integrating out the heavy fermions $\xi_n$, and keeping only the leading terms in a derivative expansion gives:
\ba
{\cal L}_{\rm on-shell}&=&-\frac{1}{2}(HL)^2\sum_{m_n\lessgtr0}\frac{y_n^2}{m_n}+{\rm hc}.
\ea
Let us consider the contribution from the low-lying modes first. In this case the sum includes both the positive and negative tower up to $m_{\rm max}<x^{\rm special}$. After some algebra we find:
\ba\label{F1}
{\cal L}_{\rm on-shell}&=&-\frac{1}{2}(HL)^2\sum^{m_{\rm max}}\frac{y_n^2}{m_n}+{\rm hc}\\\no
&=&-\frac{1}{2}(HL)^2\frac{\lambda^2_5}{dR}\left(\frac{2c_L-1}{1-\Omega^{1-2c_L}}\right)\Omega^{2+2c_N-2c_L}F(c_N)+{\rm hc},
\ea
with
\ba\label{F}
F(c_N)\equiv\frac{4^{c_N}\pi}{\Gamma^2(\nu+1)}\sum_{n}^{n_{\rm max}}\frac{1}{|x_n^0|^{2+2c_N}}\left[4c_N\left(\frac{2}{\pi|x^0_n|}\frac{1}{J_{\nu+1}^2(|x_n^0|)}\right)-2\frac{Y_\nu(|x_n^0|)}{J'_\nu(|x_n^0|)}\right].
\ea
In this expression, the Bessel functions are all evaluated at the zeros $x_n^0$ of $J_{\nu=-c_N-1/2}$. Rather than presenting the details of this computation, it is more instructive to reproduce an approximate expression valid for $n\gg1$:
\begin{adjustwidth}{-1cm}{}
\ba\label{eq:[]}
&&\sum^{m_{\rm max}}\frac{y_n^2}{m_n}\\\no
&\to&\frac{\lambda^2_5}{R^2}\left(\frac{2c_L-1}{1-\Omega^{1-2c_L}}\right)\Omega^{1-2c_L}R'\sum_{n=0}^{n_{\rm max}}\frac{1}{|x_n|}\left(\frac{1-2c_N\frac{\delta_n}{|x_n|}}{1+\frac{\delta_n}{|x_n|}}+\frac{1+2c_N\frac{\delta_n}{|x_n|}}{-1+\frac{\delta_n}{|x_n|}}\right)\\\no
&=&\frac{\lambda^2_5}{R^2}\left(\frac{2c_L-1}{1-\Omega^{1-2c_L}}\right)\Omega^{1-2c_L}R'\sum_{n=0}^{n_{\rm max}}(4c_N+2)\left(-\frac{\delta_n}{|x_n|^2}\right)\left(1+{\cal O}\left(\frac{\delta_n}{|x_n|}\right)\right)\\\no
&=&\frac{\lambda^2_5}{dR}\left(\frac{2c_L-1}{1-\Omega^{1-2c_L}}\right)\Omega^{2+2c_N-2c_L}\left[\frac{4^{c_N}\pi}{\Gamma^2(-c_N+1/2)}\sum_{n=0}^{n_{\rm max}}\frac{(4c_N+2)}{|[n+\frac{1}{2}(1-c_N)]\pi|^{2+2c_N}}\right]\left(1+{\cal O}\left(\frac{\delta_n}{|x_n|}\right)\right).
\ea
\end{adjustwidth}
One can verify that $F(c_N)$ consistently reduces to the quantity in the square bracket in this limit. $F(c_N)$ is a sole function of $c_N$. It is plotted in Figure~\ref{fig:agree} for various values of $n_{\rm max}$.

\end{document}